\documentclass[twocolumn, twocolappendix]{aastex701}
\usepackage{float}
\usepackage{tabularx}
\usepackage{caption}
\usepackage{subcaption}
\usepackage{makecell}
\usepackage{amsmath} %added by nisrine
\usepackage{graphicx}
\usepackage{gensymb}
\usepackage{float}
\usepackage{verbatim}
\usepackage{makecell}
\usepackage{multirow}
\usepackage{appendix}

\defcitealias{2023NatAs...7..986B}{J.S. Bright \& L. Rhodes et al. 2023}
\defcitealias{2019MNRAS.486.2721B}{J.S. Bright et al. 2019}
\defcitealias{2020MNRAS.496.3326R}{L. Rhodes et al. 2020}
\defcitealias{2024MNRAS.533.4435R}{L. Rhodes et al. 2024}
\defcitealias{1997Natur.389..261F}{D. Frail et al. 1997}
\defcitealias{2004ApJ...609L...1T}{G.B. Taylor et al. 2004}
\defcitealias{2005ApJ...622..986T}{G.B. Taylor et al. 2005}
\defcitealias{2007ApJ...664..411P}{Y.M. Philstr\"{o}m et al. 2007}
\defcitealias{2008ApJ...683..924C}{P. Chandra et al. 2008}
\defcitealias{2012ApJ...759....4M}{R.A. Mesler et al. 2012}
\defcitealias{2022MNRAS.513.1895R}{L. Rhodes et al. 2022}
\defcitealias{2023MNRAS.523.4992A}{G.E. Anderson et al. 2023}
\defcitealias{2024A&A...690A..74G}{S. Giarratana et al. 2024}
\defcitealias{2019ApJ...870...67A}{K.D. Alexander et al. 2019}

\begin{document}

\title{The second Arcminute Microkelvin Imager -- Large Array Gamma-ray burst radio afterglow catalog}

%% A significant change from AASTeX v6+ is in the author blocks. Now an email
%% address is required for each author. This means that each author requires
%% at least one of the following:
\author[0000-0003-2705-4941]{Lauren Rhodes}
\email[show]{lauren.rhodes@mcgill.ca}
\affiliation{Trottier Space Institute at McGill, 3550 Rue University, Montreal, Quebec H3A 2A7, Canada}
\affiliation{Department of Physics, McGill University, 3600 Rue University, Montreal, Quebec H3A 2T8, Canada}
\author{Dinah Ibrahim}
\email{dinah.ibrahim@mail.mcgill.ca}
\affiliation{Trottier Space Institute at McGill, 3550 Rue University, Montreal, Quebec H3A 2A7, Canada}
\affiliation{Department of Physics, McGill University, 3600 Rue University, Montreal, Quebec H3A 2T8, Canada}
\author[]{Nisrine Sqalli}
\email{nisrine.sqalli@mail.mcgill.ca}
\affiliation{Trottier Space Institute at McGill, 3550 Rue University, Montreal, Quebec H3A 2A7, Canada}
\affiliation{Department of Physics, McGill University, 3600 Rue University, Montreal, Quebec H3A 2T8, Canada}
\author[0000-0001-6544-8007]{Gemma E. Anderson}
\email{gemma.anderson@csiro.au}
\affiliation{Australia Telescope National Facility, CSIRO, Space and Astronomy, PO Box 1130, Bentley, WA  6151, Australia}
\affiliation{Sydney Institute for Astronomy, School of Physics, The University of Sydney, NSW 2006, Australia}
\author{Rob Fender}
\email{rob.fender@physics.ox.ac.uk}
\affiliation{Astrophysics, Department of Physics, University of Oxford, Keble Road, Oxford OX1 3RH, UK}
\affiliation{Department of Astronomy, University of Cape Town, Private Bag X3, Rondebosch 7701, South Africa}
\author[0000-0001-9149-6707]{Alexander J. van der Horst}
\email{ajvanderhorst@email.gwu.edu}
\affiliation{Department of Physics, the George Washington University, 725 21st Street NW, Washington, DC 20052, USA}
\affiliation{Astronomy, Physics and Statistics Institute of Sciences (APSIS), 725 21st Street NW, Washington, DC 20052, USA}
\author[0000-0002-7735-5796]{Joe S. Bright}
\email{jbright@seti.org}
\affiliation{Astrophysics, Department of Physics, University of Oxford, Keble Road, Oxford OX1 3RH, UK}
\affiliation{SETI Institute, 339 Bernardo Ave., Suite 200 Mountain View, CA 94043, USA}
\affiliation{Breakthrough Listen, University of Oxford, Department of Physics, Denys Wilkinson Building, Keble Road, Oxford, OX1 3RH, UK}
\author[0000-0003-0764-0687]{Andrew K. Hughes}
\email{andrew.hughes@physics.ox.ac.uk}
\affiliation{Astrophysics, Department of Physics, University of Oxford, Keble Road, Oxford OX1 3RH, UK}
\author[0000-0001-7361-0246]{David R.A. Williams-Baldwin}
\email{david.williams-7@manchester.ac.uk}
\affiliation{Jodrell Bank Centre for Astrophysics, School of Physics and Astronomy, The University of Manchester, Manchester, M13 9PL,
UK}
\author[0000-0003-3189-9998]{David A.\ Green}
\email{dag@mrao.cam.ac.uk}
\affiliation{Cavendish Laboratory, University of Cambridge, J.~J.\ Thomson Ave., Cambridge, CB3 0US, UK}
\author[0000-0001-6803-2138]{Daryl Haggard}
\email{daryl.haggard@mcgill.ca}
\affiliation{Trottier Space Institute at McGill, 3550 Rue University, Montreal, Quebec H3A 2A7, Canada}
\affiliation{Department of Physics, McGill University, 3600 Rue University, Montreal, Quebec H3A 2T8, Canada}
\author[0000-0002-5936-1156]{Assaf Horesh}
\email{assafh@mail.huji.ac.il}
\affiliation{Racah Institute of Physics, The Hebrew University of Jerusalem, Jerusalem 91904, Israel}
\author{Kunal P. Mooley}
\email{kmooley@iitk.ac.in}
\affiliation{Indian Institute Of Technology Kanpur, Kanpur, Uttar Pradesh 208016, India}
\affiliation{Caltech, 1200 E. California Blvd. MC 249-17, Pasadena, CA 91125, USA}
\author[0000-0002-6255-8240]{Yvette Perrott}
\email{yvette.perrott@vuw.ac.nz}
\affiliation{School of Chemical and Physical Sciences, Victoria University of Wellington, PO Box 600, Wellington 6140, New Zealand}
\author{Paul Scott}
\email{paul@mrao.cam.ac.uk}
\affiliation{Cavendish Laboratory, University of Cambridge, J.~J.\ Thomson Ave., Cambridge, CB3 0US, UK}
\author{Tim Staley}
\email{tim@timstaley.co.uk}
\affiliation{Astrophysics, Department of Physics, University of Oxford, Keble Road, Oxford OX1 3RH, UK}
\author{David J.\ Titterington}
\email{djt@mrao.cam.ac.uk}
\affiliation{Cavendish Laboratory, University of Cambridge, J.~J.\ Thomson Ave., Cambridge, CB3 0US, UK}

\begin{abstract}
Radio observations of gamma-ray burst afterglows provide insight in to the different emitting regions within the jet, its hydrodynamics and burst environment. In this paper, we present the second iteration of the Arcminute Microkelvin Imager -- Large Array (AMI--LA) Gamma-ray burst (GRB) radio afterglow catalog. The catalog consists of 1035 observations of 210 bursts. Our observations range from 0.04 to 900\,days post-burst. We detect radio emission associated with 41 events with flux densities spanning 0.1-40\,mJy, and present a detailed analysis of six events whose light curves have not been published elsewhere. In our individual afterglow analyses, we find that our radio counterparts show evidence of reverse shock emission 60\% of the time. Two of the detected afterglows show evidence of scintillation which we use to place source size limits at tens of days. Of the whole catalog, 50 events have redshift measurements which we use to find the radio afterglow luminosity distribution for our sample. Our luminosity distribution spans five orders of magnitude, a larger range than seen in optical and X-ray bands, indicative of the strong dependence of the radio counterparts on the GRB's physical parameters. 
\end{abstract}
%strong variety from event and event including scintillation, strong reverse shock components and jet breaks. 


%% From the front matter, we move on to the body of the paper.
%% Sections are demarcated by \section and \subsection, respectively.
%% Observe the use of the LaTeX \label
%% command after the \subsection to give a symbolic KEY to the
%% subsection for cross-referencing in a \ref command.
%% You can use LaTeX's \ref and \label commands to keep track of
%% cross-references to sections, equations, tables, and figures.
%% That way, if you change the order of any elements, LaTeX will
%% automatically renumber them.

\section{Introduction}

Gamma-ray bursts (GRBs) are brief ($\sim$seconds-long) luminous flashes of high energy radiation ($\sim$MeV energies) emitted when highly relativistic jets are launched in cataclysmic events such as massive stellar explosions or binary neutron star mergers \citep{1992MNRAS.257P..29M,1995ApJ...455L.143S}. The GRB's progenitor system can usually be inferred from the duration of the prompt emission, i.e. the length of the GRB, a burst longer than two seconds usually comes from a collapsing massive star whereas a short burst (lasting less than two seconds) is typically associated with binary neutron star mergers \citep{1993ApJ...413L.101K,1993ApJ...405..273W,2017ApJ...848L..13A}. These two classes of events are referred to as long and short GRBs, respectively. The prompt emission is followed by an \textit{afterglow} which is produced when the relativistic jet interacts with the circumburst environment forming shocks \citep{1976PhFl...19.1130B,1997MNRAS.288L..51W,1998ApJ...497L..17S}. The shocks sweep up material (electrons) in the circumburst environment and accelerate them into a power law distribution of energies ($N(E)dE \propto E^{-p}dE$, with $2 \lesssim p \lesssim 3$) that cool emitting synchrotron radiation. The synchrotron emission dominates the afterglow between radio and X-ray energies. At higher frequencies, it is possible that synchrotron self-Compton/external inverse Compton emission may also be detected \citep{2019Natur.575..455M,2021Sci...372.1081H,2023Sci...380.1390L}.

The afterglow emission arises from a superposition of multiple synchrotron spectra (i.e different shock regions). The dominant emitting region at most wavelengths is the \textit{forward shock}, a shock that forms between the jet and the circumburst environment \citep{1998ApJ...497L..17S}. While the forward shock moves outwards (away from the launch site) there is a corresponding \textit{reverse shock} moving in the opposite direction \citep[in the rest frame of the jet, ][]{1999ApJ...520..641S,1999MNRAS.306L..39M,2003ApJ...582L..75K}. The strength of the reverse shock is dependent on the jet's physical parameters including the jet Lorentz factor, magnetism and shock depth. 

Each synchrotron spectrum that evolves as the jet expands and decelerates. A given synchrotron spectrum is characterized by three break frequencies: the self-absorption frequency, $\nu_{\rm{sa}}$, the minimum energy electrons break, $\nu_{\rm{m}}$, and the cooling frequency, $\nu_{\rm{c}}$. The peak of the spectrum and the peak flux density are dictated by whichever of $\nu_{\rm{sa}}$ or $\nu_{\rm{m}}$ is at a higher frequency. The location of each break frequency can be constrained with multi-frequency observations across the radio to X-ray range or multi-epoch observations in a smaller number of observing bands. Observing at a single frequency across multiple epochs, a light curve will look like a power law until a frequency break moves through the band and causes the the light curve to change shape. Frequency breaks thus cause chromatic light curve breaks. 

Locating the position of each frequency break and how they move with time allows one to extract physics parameters of the afterglow: microphysical parameters, the kinetic energy, the circumburst density/profile ($\rho \propto r^{-k}$) and jet geometry \citep{1998ApJ...497L..17S,2000ApJ...536..195C,Granot_2002}. However, the difference frequency breaks are all dependent on the aforementioned physical parameters in different ways, therefore the observer needs to locate all the breaks to resolve degeneracies in the afterglow modeling. Given that both $\nu_{\rm{sa}}$ and $\nu_{\rm{m}}$ are moving through the radio band, without radio observations, afterglow parameter estimation is not possible.

In addition to light curve changes caused by spectral evolution, the light curve can also vary due to geometric effects such as a \textit{jet break}, which occurs when the jet has decelerated sufficiently such that the observer is able to see the edges of the jet and not just a small cone of emission due to relativistic beaming \citep{1998ApJ...503..314P,1999ApJ...519L..17S}. Geometric effects, such as jet breaks, are distinguishable from other light curve changes achromatically. However, more detailed studies of both the theory and observations around jet break times have demonstrated that geometric effects may still have some chromatic component \citep[e.g.][]{2011MNRAS.410.2016V, 2026ApJ...997L...1L}, which can further complicate any interpretation. 

The most simple jet geometry is a \textit{top hat} jet which has some constant kinetic energy with opening angle out to the jet opening angle at which the energy is zero \citep[e.g.][]{2005ApJ...631.1022G,2018ApJ...863...32D,2018MNRAS.481.2711G}. More complex jet structures, geometries and outflow components such as cocoons have also been used to explain light curve and spectral features of afterglows \citep[e.g.][]{2022MNRAS.513.1895R, 2024MNRAS.533.4435R}. We do not consider such geometries in this work.

On shorter timescales (hours to $\sim$day), variability can occur as a result of \textit{scintillation} \citep{1997NewA....2..449G, 1998MNRAS.294..307W}. Scintillation occurs as a result of material along the line of sight (most likely in the Milky Way), refracting and diffracting the radio waves that originate from the GRB afterglow such that the radio emission we detect modulates with frequency and time \citep[e.g.][]{1997Natur.389..261F,2014MNRAS.444.3151V,2023MNRAS.523.4992A}. The effects of scintillation are only visible because the source of the radio emission is sufficiently compact. Therefore, by observing scintillation, we can place limits on the emitting region source size.

%With sufficiently detailed radio observations over multiple epochs, it is also possible to extract the circumburst density profile and constrain the geometry of the jet. 

Most works studying GRB afterglows focus on single events. However, there are also a number of catalog efforts that often concentrate at optical or X-ray wavelengths. To date, there have been two radio catalogs published \citep{2012ApJ...746..156C,2018MNRAS.473.1512A}. \citet{2012ApJ...746..156C} published observations of 304 GRBs taken across a range of frequencies, over a span of 14\,years with the Karl G. Jansky Very Large Array, both prior and post upgrade, and the Australia Telescope Compact Array (ATCA). They found that 31\% of GRBs had a radio detection and that radio emission was more likely to be detected from events with higher gamma-ray fluence. The authors note that in the pre-\textit{Swift} era, the number of GRB detections was sufficiently low such that nearly all events were observed by the VLA. The follow up rate in the \textit{Swift} era is far lower due to the higher discovery rate and therefore follow up has been more selective potentially inducing biases in which events were observed.  %The most likely origin of the high detection fraction in \citet{2012ApJ...746..156C} is the nature of the programs running on the VLA. In the pre

\citet{2018MNRAS.473.1512A} published the first Arcminute Microkelvin Imager -- Large Array (AMI--LA) catalog for GRB radio afterglows, consisting of 139 GRBs observed over 3 years. We refer to this catalog as AMI-GRB-1. All observations were conducted at 15.7\,GHz. \citet{2018MNRAS.473.1512A} found a detection fraction of 15\% down to a sensitivity limit of $\sim0.2$\,mJy. Of the 139 events reported in \citet{2018MNRAS.473.1512A}, 132 GRBs were observed as part of the AMI--LA Rapid Response Mode (ALARRM) program which enabled data collection as early as two-minutes post-burst resulting in some of the earliest radio afterglow detections \citep{2013MNRAS.428.3114S,2014MNRAS.440.2059A}. The rapid response system enabled detections of afterglow emission when they are still in a reverse-shock dominated regime. Using the early time radio detections/limits 
\citet{2018MNRAS.473.1512A} places limits on the bulk Lorentz factor of around 100 at the time of the radio observations. For afterglows with redshifts, the authors found that the radio luminosities spanned 10$^{29-32}$\,erg/s. 

Similar rapid response systems have been implemented at e.g. LOFAR \citep{2024MNRAS.534.2592R}, MWA \citep{2015ApJ...814L..25K,2019PASA...36...46H,2021PASA...38...26A}, ATCA \citep{2021MNRAS.503.4372A, 2024ApJ...975L..13A}, the Sub-millimeter Array \citep[SMA, e.g. ][]{2026arXiv260414297K}, and the VLA (accepted proposal VLA/26B-082, PI: van der Horst), thus increasing the sensitivity and frequency coverage of rapid response systems. These existing faculties will lay the ground work for rapid response triggering on the Square Kilometer Array \citep{2026arXiv260703024A}.%\textcolor{red}{Gemma's SKA chapter}.

%Both AMI—LA and ATCA have both demonstrated the important of rapid response observations in past works, e.g. GRB 130427A or short GRB 231117 \citep{2014MNRAS.440.2059A, 2025ApJ...994....5A}. %However, observations within a day post-discovery also need to be complemented by subsequent observations that logarithmically scaled in order to truly rule out radio the presence of radio emission. 

In this work, we present the second AMI--LA radio catalog which consists of observations of 210 GRBs over a ten year period. We refer to this new catalog as AMI-GRB-2. The vast majority of the GRBs observed are \textit{long} GRBs, with the exception of four events: GRBs 180715A, 180718A, 190610A and 210627A, which are \textit{short} GRBs. Therefore any event with the title \textit{GRB} is a long GRB (most likely from a collapsar) and we refer to short GRBs (likely binary neutron star mergers) as short GRBs. In Section \ref{sec:methodology}, we describe the data reduction method for all AMI--LA observations along with a brief description of how the multi-wavelength counterpart data was collected for all the bursts presented here. We have split our analysis into consideration of the individual events that make up the catalog: analysis of the population as a whole in Section \ref{sub:population} and detailed analysis of individual events in Section \ref{sub:individual}. In Sections \ref{sec:disc} and \ref{sec:conc}, we discuss the implications of our findings in the context of other radio GRB catalogs as the known GRB population as a whole and present our conclusions. 

We adopt a $\Lambda$CDM cosmology throughout this paper with H\textsubscript{0}= (67.8 $\pm$ 0.9) km/s/Mpc, a matter density parameter $\Omega_m$ = 0.308 $\pm$ 0.012 \citep{2016A&A...594A..13P}

\section{Methodology} \label{sec:methodology}

\subsection{Observations: AMI--LA}\label{sub:observations}

The Arcminute Microkelvin Imager -- Large Array (AMI--LA) is an eight-dish interferometer based at the Mullard Radio Astronomy Observatory in Cambridge, UK \citep{2008MNRAS.391.1545Z}. The array observes at a central frequency 15.5\,GHz with a bandwidth of 5\,GHz, is split into 8 spectral windows. The maximum baseline is 110\,meters corresponding to an angular resolution of about half an arcminute.

In this paper we consider all bursts observed with AMI--LA between January 2016 and January 2025. In total we consider 210 GRBs. Of the 210 GRBs presented in this catalog, 107 of them were observed automatically through the ALARRM program \citep{2013MNRAS.428.3114S, 2018MNRAS.473.1512A}. The ALARRM program runs by accepting notifications via a VOEvent from the \textit{Swift} - Burst Alert Telescope when it detects a GRB \citep{2016arXiv160603735S}. The VOEvent triggers a rapid response observation of the GRB if it is visible to AMI--LA. Not all observations were obtained through the ALARRM system. AMI--LA did not observe for several months during the COVID-19 pandemic, and the ALARRM system was switch off. All subsequent observations were manually scheduled thereby increasing the time between the first observation and the initial trigger.

The exact timings of our observing strategy has changed slightly over the period of observations we present here. The ideal scenario would be logarithmically spaced observations starting as soon as possible post burst with at least three observations to allow for the detection of later onset radio emission. In reality, there are many external factors that inhibit a perfect observing strategy such as scheduling conflicts or loss of antennas. We note that of the GRBs within in this catalog, 73 were observed only once. 101 events were observed between 2 and 6 times with observations spaced logarithmically. The remaining events had higher observation numbers as a result of longer detections and so were observed until the source were not longer detectable. The standard duration for AMI--LA manually scheduled observation was 4\,hours and for ALARRM observations: 2\,hours.

\begin{figure*}
    \centering
    \includegraphics[width=\linewidth]{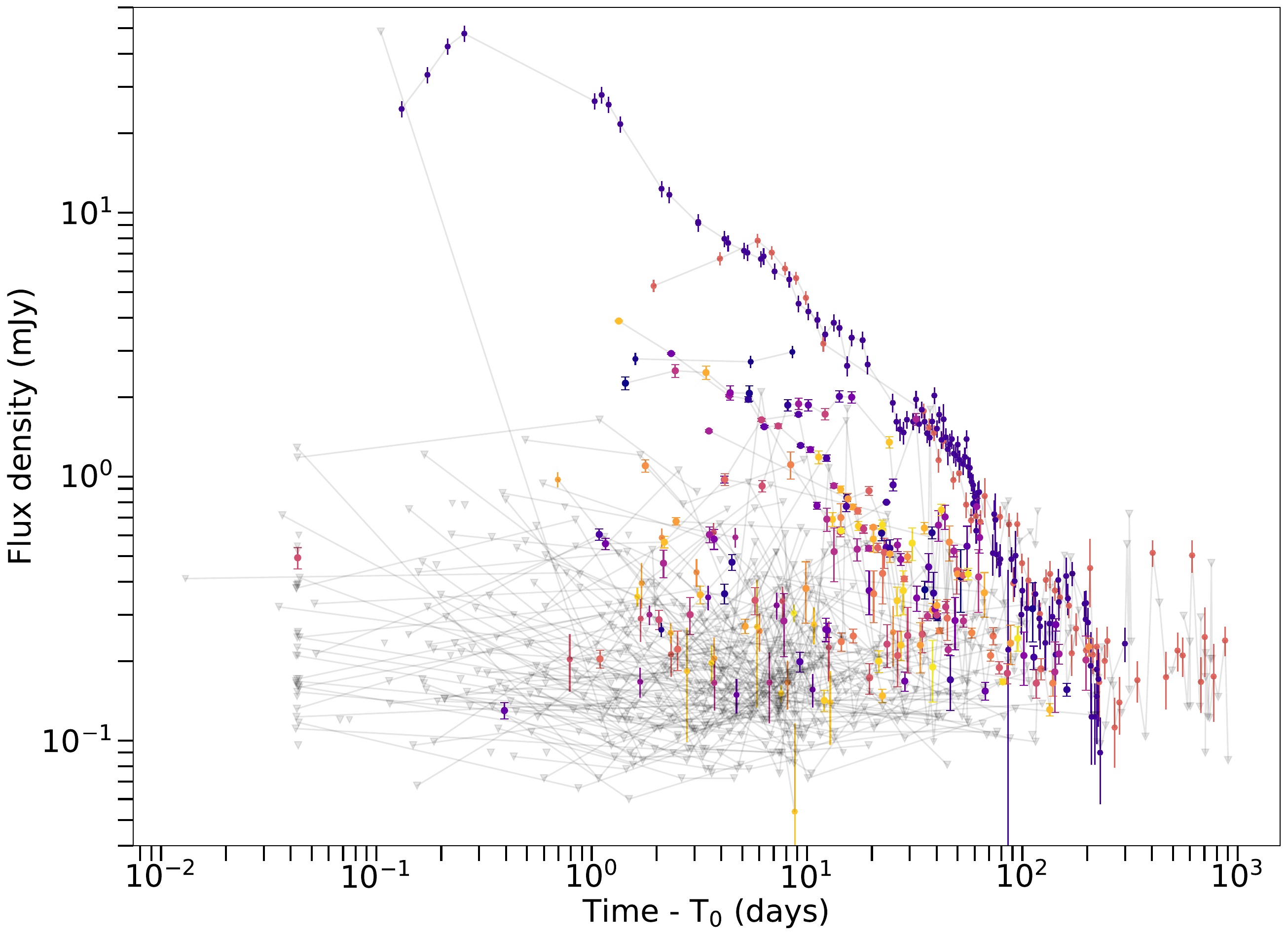}
    \caption{A compilation of all the GRBs observed between January 2016 and January 2025 (i.e. AMI-GRB-2). The downwards facing triangles represent 3$\sigma$ upper limits and the lines connect individual events. }
    \label{fig:full_cat}
\end{figure*}

All AMI--LA data is processed using a custom software \textsc{reduce\_dc} which converts the raw data into \textsc{fits} files, flags for radio frequency interference, antenna shadowing and known instrumental problems, it performs amplitude and complex gain calibration. All cleaning and deconvolution is performed iteratively using the task \textit{clean} within \textsc{casa} (Version 4.7.0). Once the images are made, we extracted the target flux densities using \textit{imfit}. In Figure \ref{fig:full_cat}, we show the full AMI-GRB-2 catalog, including events where radio emission was detected at the position reported by GRB monitors or afterglow discoveries at other wavelengths.

\subsection{Optical and X-ray Data}\label{sub:op_x_data}

This work focuses on the importance of radio observation in informing our understanding of GRB afterglows. However there are cases where optical and X-ray data have been useful in contextualizing our radio observations for specific events. In these cases we have taken optical data from published papers and the \textit{General Coordinates Network}\footnote{https://gcn.nasa.gov}, and X-ray data from the \textit{Swift} Burst Analyser website\footnote{https://www.swift.ac.uk/burst\_analyser/}. From the \textit{Swift} Burst Analyser, we use the 10\,keV flux density light curve and the photon index derived from spectral fitting. We chose to use flux density over flux or counts because this allows us to swap between flux density-time and flux density-frequency space for afterglow analysis more easily. We use the flux data in one case, GRB 210610B, where we find substantial differences between the flux density light curve and the \textit{Swift}-XRT catalog entry.

%\begin{figure*}
%    \centering
%    \includegraphics[width=0.8\linewidth]{total_sample.pdf}
%    \caption{The full 10 year AMI--LA GRB dataset. All detections of radio emission with coordinates consistent with a GRB are marked by purple circles with errorbars. The downwards facing triangles represent 3$\sigma$ upper limits. }
    %\label{fig:full_cat}
%\end{figure*}

\section{Results: the full sample}\label{sub:population}

Our sample consists of 210 GRBs observed, including 41 detected events, between January 2016 and 2025, with a total of 1035 flux density measurements at 15.5 GHz. Of those 1035 flux density measurements, 386 are detections and 649 are upper-limits. The data shown in this section is all available online as a downloadable catalog\footnote{https://laurenrhodes.github.io/index\_catalog.html}.

Figure \ref{fig:full_cat} shows the AMI-GRB-2 catalog including the light curves for the 41 systems where we detected radio emission. Our observations span from 0.04--900\,days post-burst. Our flux densities measurements range from tens mJy down to the noise floor of $\sim0.1$\,mJy. The time stamp for each data point is the central time of each observations.

At less than one day post-burst, we have mostly non-detections, and our upper limits have a large spread extending between 0.1 and 1\,mJy. There is also a vertical line of limits (and a single detection) at around 1\,hour post-burst. For sources that are visible to AMI--LA at the time of reporting, ALARRM enables AMI--LA to slew immediately to the position of the burst and performs a shorter two-hour observation (standard observations are four-hours long). Therefore, when we plot the light curves, any burst that was immediately observed will have a $t_{\rm{mid}} - t_0$ of one hour.

%Figure \ref{fig:full_cat_dets} shows that the flux densities of first radio observations span from a few mJy down to our noise limit of $\sim0.1$\,mJy, with an average flux density of 0.9$\pm$0.8\,mJy (1$\sigma$ uncertainties). We also show the flux density distribution of the whole catalog in Figure \ref{fig: Flux Density Distribution}. 
%Figure \ref{fig: Flux Density Distribution} shows the flux density distribution of the 3-sigma upper-limits as well as the detections first for all the observations in the data set (Figure \ref{fig: Flux Density All}), then for observations conducted 10 days or less after the initial burst (Figure \ref{fig: Flux Density 10 days}). %When considering only early-time observations ($t \leq 10$ days), the number of detections remains more or less flat across the entire range of flux densities, while the number of non-detections peaks in the $100 - 150 \mu Jy$ range then reduces as the flux density increases. Figure \ref{fig: Flux Density 10 days} shows that detection rate distribution is flat with flux density at early times (within the first 10\,days, i.e. when the afterglow is being discovered), down to our detection limit. 

\begin{figure}[!ht]
\begin{subfigure}{\linewidth}
  \includegraphics[width=0.95\linewidth]{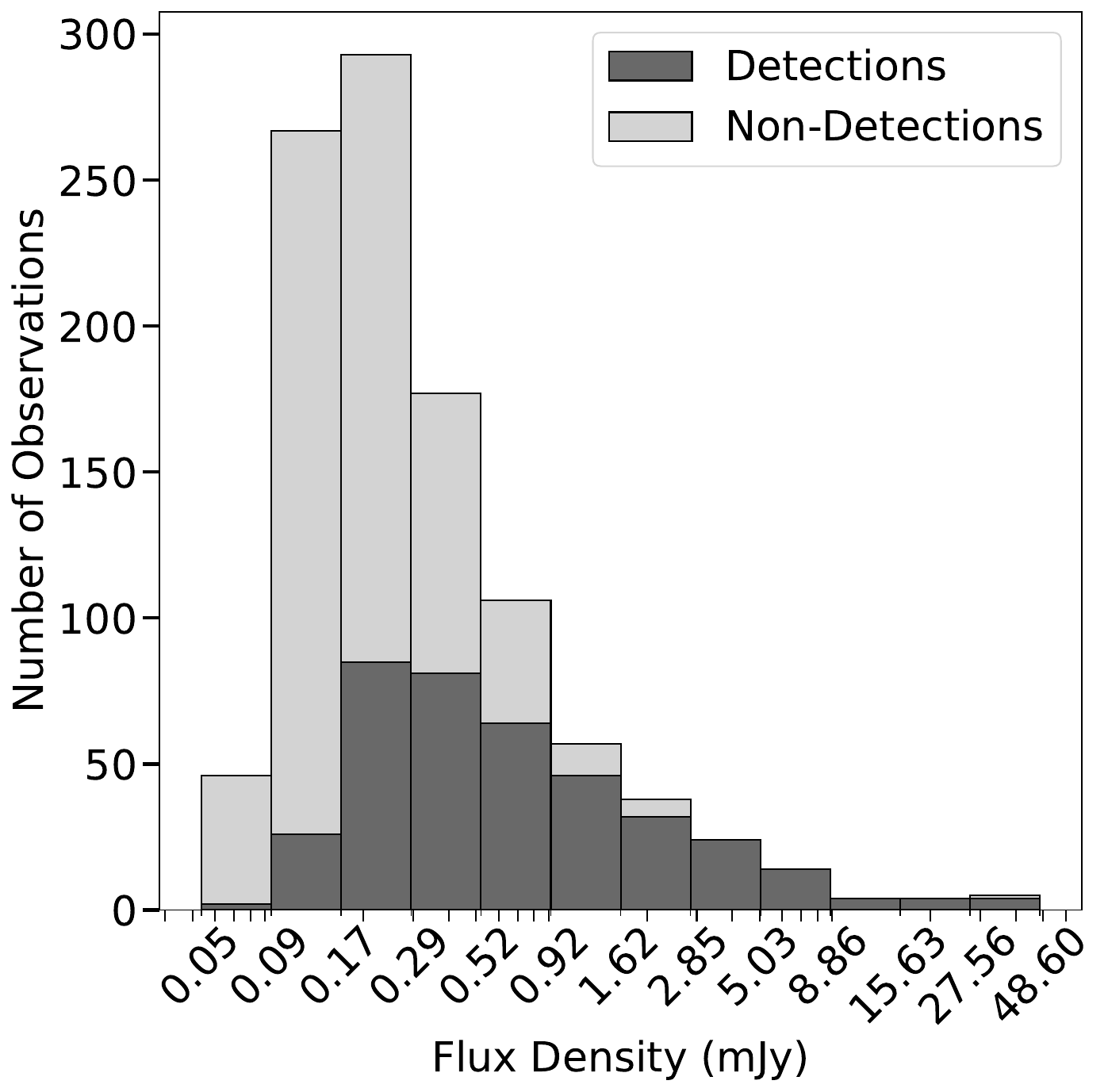}
  \caption{Distribution of entire population data.}
  \label{fig: Flux Density All}
\end{subfigure}\hfill % <-- "\hfill"
~ % optional tilde b/t figures for readability.
  % this solution will not work w/ empty lines b/t subfigures
\begin{subfigure}{\linewidth}
  \includegraphics[width=0.95\linewidth]{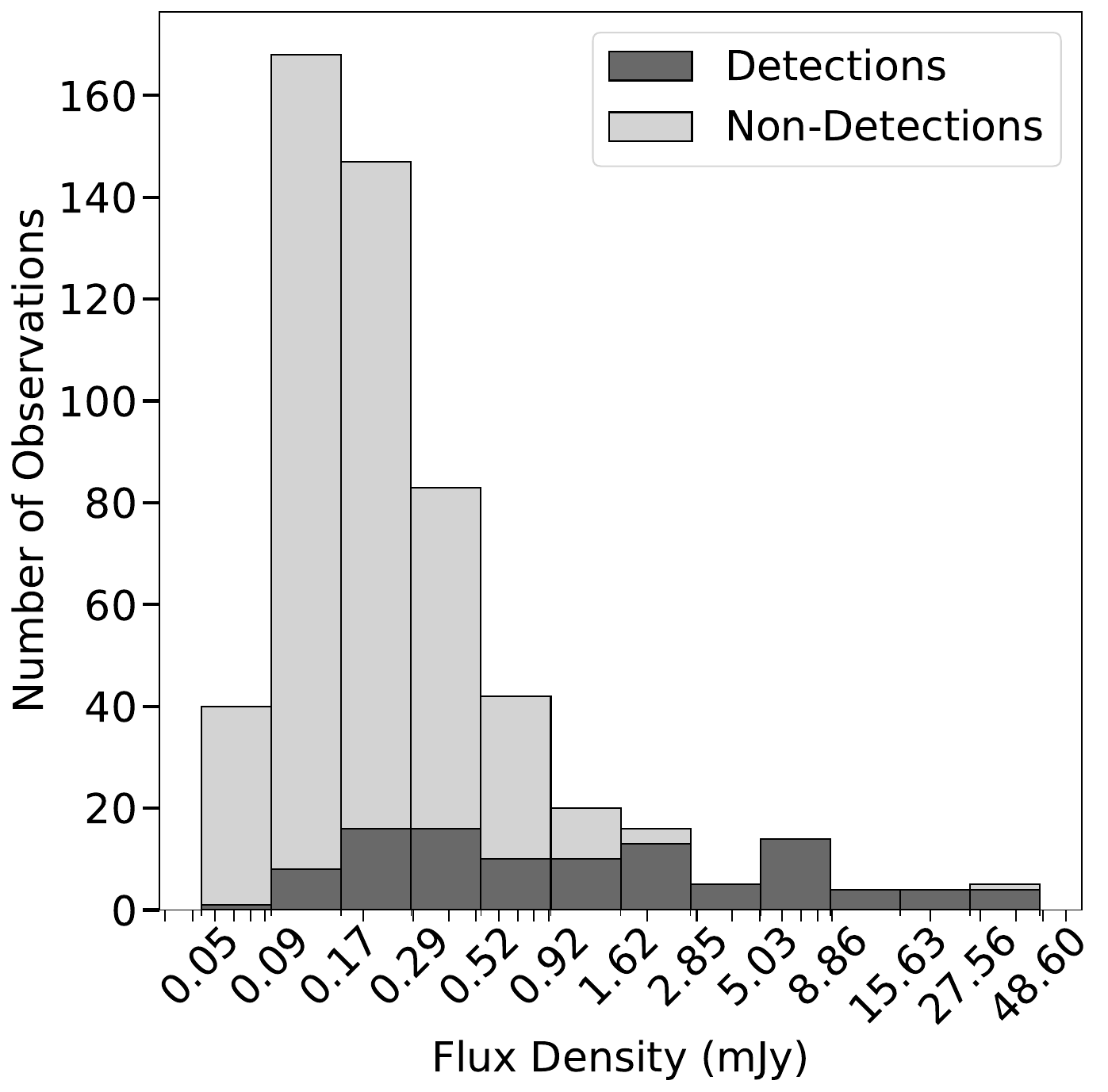}
  \caption{Distribution of data with $t \le 10$ days.}
  \label{fig: Flux Density 10 days}
\end{subfigure}
\caption{The flux density distribution plotted as stacked histograms for (top) all observations within AMI-GRB-2 and (bottom) of the observations between 0 and 10 days since burst. The light gray bars represent the distribution for the 3$\sigma$ upper limits, and the dark gray bars represent the distribution of the detections. The early-time distribution of detections is flat across the range of flux densities. The typical 3$\sigma$ upper limit of any AMI-LA observation is 0.1-0.2\,mJy.}
\label{fig: Flux Density Distribution}
\end{figure}

In Figure \ref{fig: Flux Density Distribution}, we collapse the time axis of Figure \ref{fig:full_cat} and show the flux density distribution of our upper limits and detections. In Figure \ref{fig: Flux Density All} shows that the flux density distribution peaks around 0.2\,mJy i.e. near the sensitivity limit of the telescope. In the lower flux density bins, we see a very slight reduction in the detection fraction. We also show the distribution for the first ten days post-burst, when we have the largest range of flux density measurements. Figure \ref{fig: Flux Density 10 days} shows a flat distribution in our detections from a few mJy down to our noise limit of $\sim0.1$\,mJy. Both panels indicate that our sample is sensitivity limited and that we do not observe the whole population. Therefore, the shape of the histogram in Figure \ref{fig: Flux Density All} is most likely dominated by bright detections of afterglow emission after 10\,days post-burst.

\begin{figure}[ht]
    \centering
    \includegraphics[width=\linewidth]{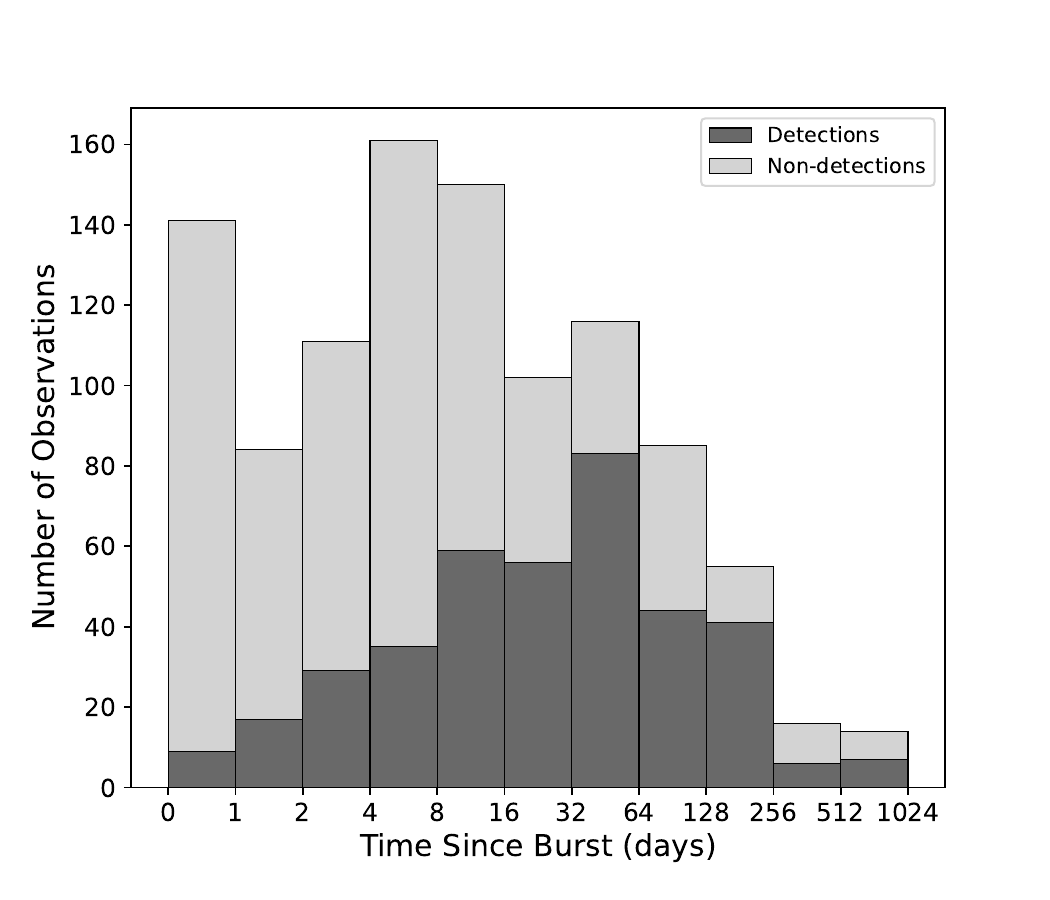}
    \caption{A stacked histogram summarizing observing times elapsed since burst for the AMI-GRB-2 catalog. The light gray and dark gray bars, respectively, represent the non-detections and detections in the data.}
    \label{fig:Time Since Burst Distribution}
\end{figure}

Figure \ref{fig:Time Since Burst Distribution} shows a stacked histogram summarizing the distribution of the data in our catalog of bursts as a function of time elapsed since the burst. It demonstrates a very low detection rate in the first 24\,hours post-burst and subsequently shows the importance of observing bursts multiple times as not all afterglows will be detected at radio frequencies (even higher frequencies like AMI—LA) in the first one or two days post-burst. There is a spike in the number of observations at less than 1\,day post-burst, almost entirely from our rapid response observations. In the subsequent time bins, we have varying numbers of observations but the number of detections increases with each time bin as we continue to observe detected events.

\begin{figure*}
    \centering
    \includegraphics[width=\linewidth]{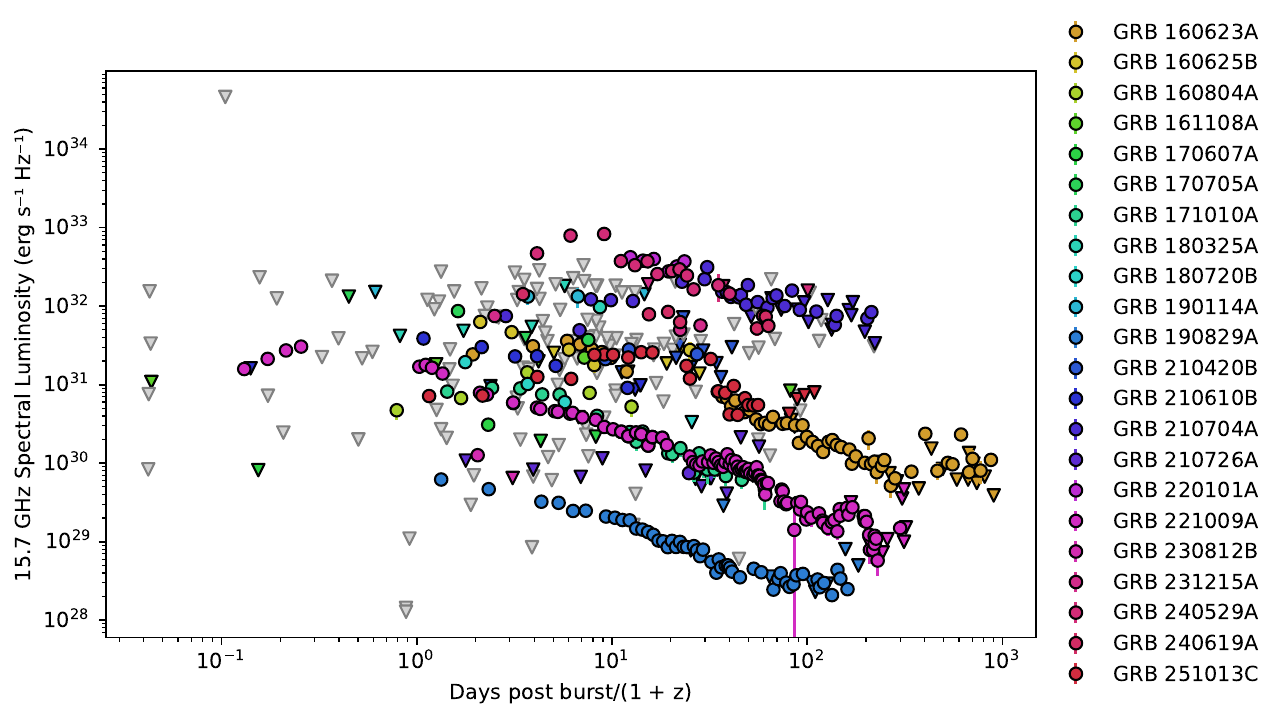}
    \caption{The \textit{k}-corrected spectral luminosity light curves of the AMI-GRB-2 catalog events included 22 detected afterglows in the rest frame. The GRBs are color-coded for distinction, with their detections represented as dots and their 3$\sigma$ upper-limits represented as inverse triangles. The GRBs with a redshift measurement which were observed but non-detected by AMI are represented by gray triangles \citep{2017GCN.21987....1T,2019GCN.25252....1R,2018GCN.22996....1V,2019GCN.25565....1V,2021GCN.30271....1P,2021GCN.30692....1Z,2021GCN.30000....1M,2021GCN.30164....1Z,2021GCN.29684....1H,2021GCN.31107....1R,2021GCN.29853....1D,2021GCN.30182....1F,2021GCN.30545....1K,2021GCN.30436....1W,2022GCN.31423....1C,2022GCN.31800....1D,2022GCN.31423....1C,2022GCN.31453....1R,2022GCN.32648....1D,2022GCN.31359....1F,2023GCN.34409....1D,2023GCN.34485....1M,2023GCN.34485....1M,2023GCN.35373....1T,2024GCN.36498....1D,2024GCN.36574....1D,2024GCN.37469....1D,2024GCN.37129....1S,2024GCN.37467....1D,2025GCN.38759....1Z,2025GCN.42227....1M}}
    \label{fig: Redshift Lightcurves}
\end{figure*}

Figure \ref{fig: Redshift Lightcurves} shows the radio luminosity light curves (colored circles) and $3\sigma$ upper limits (gray triangles) for the events in the AMI-GRB-2 catalog that also have a reported redshift. In total 53 GRBs observed with AMI--LA have reported redshifts. For 22/53 GRBs, radio emission was detected. We calculate the radio luminosity of each data point using the equation $L = 4\pi F_{\nu} D_{L}^{2} /(1+z)^{\alpha-\beta-1}$; where $\alpha = -1.5$ and $\beta = -0.75$ representative of an adiabatically expanding source producing optically thin synchrotron emission \citep{Granot_2002}. We detect afterglows between $10^{28}$ and $10^{33}$\,erg~s$^{-1}$Hz$^{-1}$, including six events below $10^{31}$\,erg~s$^{-1}$Hz$^{-1}$. The events below $10^{31}$\,erg~s$^{-1}$Hz$^{-1}$ are consistent with a population of low luminosity GRBs which are found only at lower redshifts \citep[due to sensitivity limits, ][]{2007ApJ...662.1111L,2017ApJ...850..117D,2020MNRAS.496.3326R,  2021ApJ...907...60M, 2022A&A...664A..36G}.

\begin{figure}%{L}{0.45\textwidth}
    \centering
    \includegraphics[width=0.98\linewidth ]{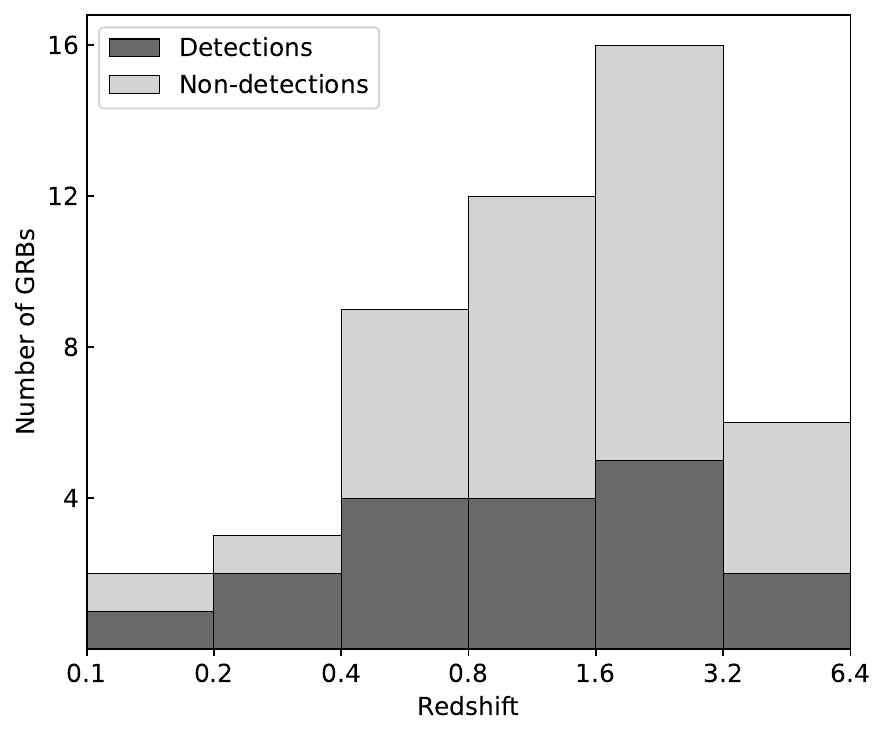}
    \caption{A stacked histogram showing the redshift distribution of the 53 GRBs from AMI-GRB-2 with redshift measurements including 22 detections.}
    \label{fig: Redshift Distribution}
\end{figure}

We also plot the distribution of redshifts for our catalog in Figure \ref{fig: Redshift Distribution} in log2 scale. The number of observations peaks in the $1.6 - 3.2$ redshift range, which contain most of the data. We find no evidence that we are preferentially detecting radio emission from afterglows in a given redshift bin. The fraction of GRBs radio-detected appears to be relatively redshift independent. We compare our observations and detections with redshift measurements to a large sample of redshifts from the \textit{Swift}-detection GRB host galaxies and find that our observations trace the redshift distribution of the \textit{Swift} GRB host galaxies sample \citep[SHOALS, ][]{2016ApJ...817....7P}. This is perhaps unsurprising given our attempts to be agnostic to afterglow properties in terms of triggering AMI--LA observations.

\section{Results: Individual events}\label{sub:individual}

We detect 41 events (shown Figure \ref{fig:full_cat}). Four of those events have already been published: GRBs 171010A, 180720B, 190829A, 221009A \citepalias{2019MNRAS.486.2721B, 2020MNRAS.496.3326R, 2023NatAs...7..986B, 2024MNRAS.533.4435R}. The non-detections from short GRB 210726A has also been published \citep{2024ApJ...970..139S}.

Here, we present in depth analysis of six events for which we have sufficient data: GRBs 160623A, 210610B, 210704A, 240529A, 240619A and 251013C (these data have not yet been published elsewhere). For each event, we characterize the afterglow light curves and spectra using the convention $F_{\nu} \propto t^{\alpha}\nu^{\beta}$. The goal of our afterglow analysis here is to determine the origin of the radio emission in terms of which branch of the synchrotron spectrum the AMI--LA observing frequency sits and whether it originates from the forward or reverse shock (or a superposition of the two). We fit phenomenological models to the AMI--LA radio data as well as public optical and/or X-ray data. Our data can be described by a combination of the following two equations. The first is 
\begin{equation}\label{eq: PL}
    y(t)= A\times t^\alpha
\end{equation}
which describes a single power law equation or for a smoothly broken power law with a single break:
\begin{equation}\label{eq:brkn_pwl}
    y(t)= A \left( \frac{1}{2}\left( \frac{t}{t_b}\right)^{-\omega\alpha_1} + \frac{1}{2}\left( \frac{t}{t_b}\right)^{-\omega\alpha_2}\right)^{-1/\omega} + B
\end{equation}
where $\omega$ is the positive smoothing parameter which is fixed to be 2. When fitting one of the above equations to our data, we are solving for all or some fraction of the amplitude \textit{A}, the temporal power index(s) $\alpha_{1, 2}$, and the break time $t_b$. 

All fits were performed using \textsc{emcee}, a Python implementation of a Monte Carlo Markov Chain sampler. We used 100 walkers, 10000 steps and discarded the first 5000. Flat priors were used at all times. The best-fit parameters are quoted as the 50th percentile of the samples in the marginalized distributions, with the lower and upper uncertainties corresponding to the 16th and 84th percentiles, respectively. For each event and waveband, we have generated a plot showing two panels. The top panel shows 100 posterior samples from our model fit. The bottom panel shows the residuals comparing the data and the model. We interpret the fits to our data using analytical predictions of the fireball model from \citet{Granot_2002, 2014MNRAS.444.3151V}.

There are also four events which contain a radio counterpart but for which we only have a single observation or do not have sufficient temporal coverage to confirm the radio counterpart is varying GRBs 220101A, GRB 220412A, 231215A \citep{2023GCN.35397....1R} and 240511A \citep{2024GCN.36489....1R}.
%150518A is all late time observations 

There are some GRBs for which observations were made but the resulting data is not included in this work: GRBs 190404B, 190510A and 190515A all had localisation regions larger than the AMI--LA primary beam making it impossible to determine if radio counterpart was present. For GRBs 180812A and 181030A we were unable to get reliable calibration solutions. GRB 171212A was classified as a possible Galactic source \citep{2017GCN.22243....1T}.

\subsection{GRB 160623A}
\label{sub:GRB160623A}

\begin{figure}[!ht]
\begin{subfigure}{\linewidth}
  \includegraphics[width=0.95\linewidth]{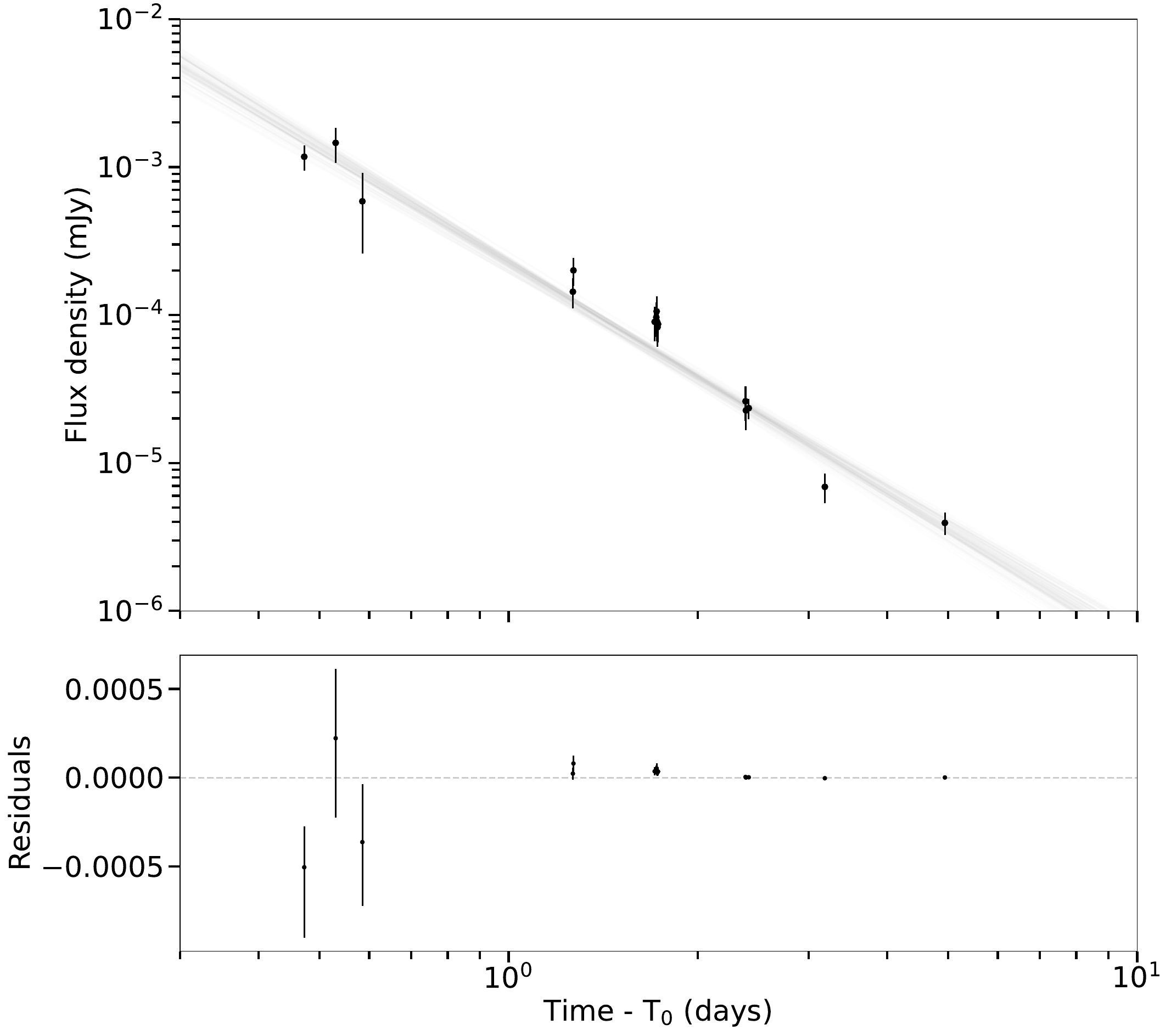}
  \caption{The \textit{Swift}-XRT light curve for GRB 160623A. The small residuals shown in the lower panel indicate that the data follows a single power law.}
  \label{fig:160623A_lc_Xray}
\end{subfigure}\hfill % <-- "\hfill"
~ % optional tilde b/t figures for readability.
  % this solution will not work w/ empty lines b/t subfigures
\begin{subfigure}{\linewidth}
  \includegraphics[width=0.9\linewidth]{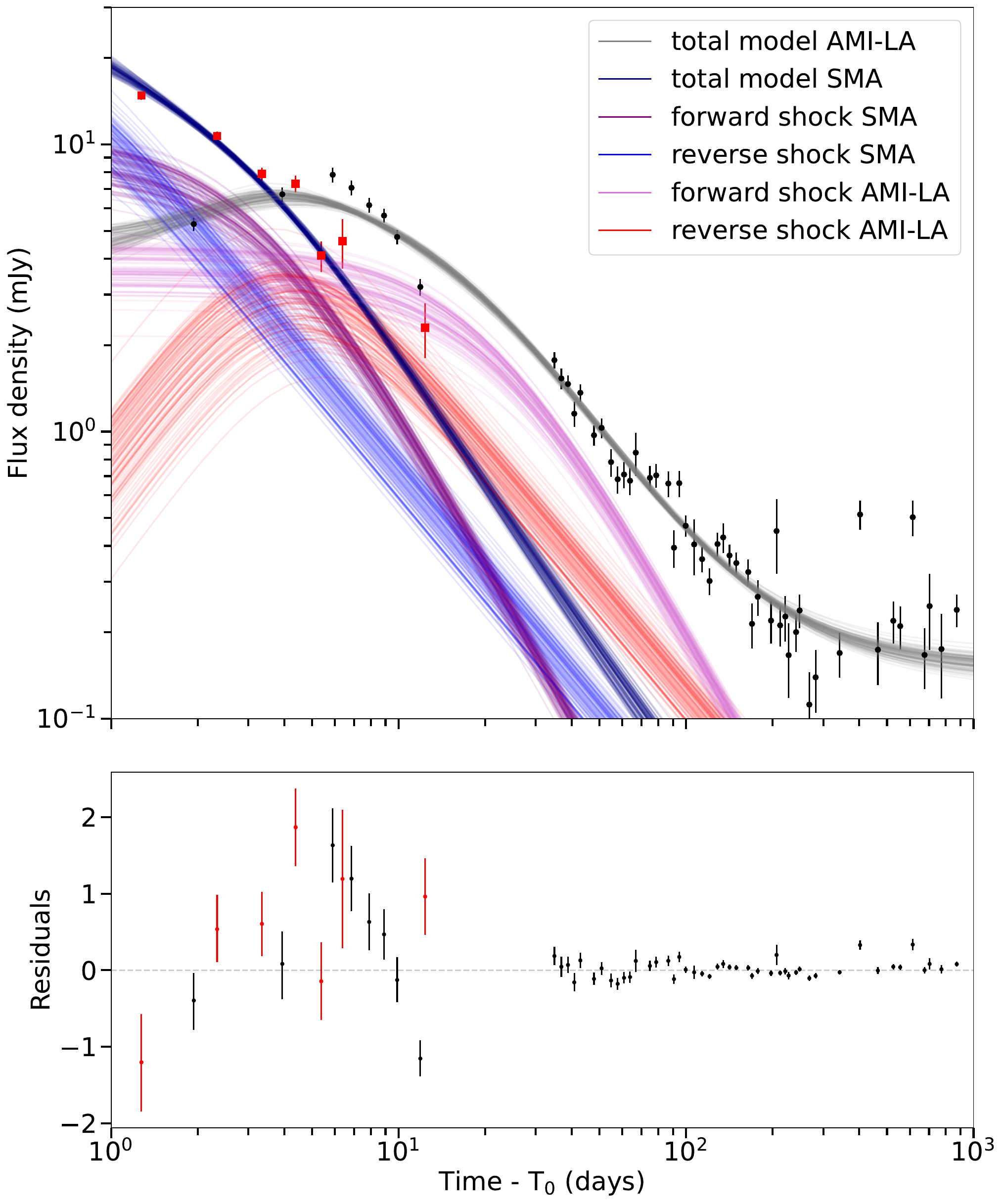}
  \caption{The AMI--LA 15.5\,GHz from our catalogue (black data points) and previously-published SMA 230\,GHz (red data points) light curve for GRB 160623A \citep{2020ApJ...891L..15C}. We fit a superposition of a forward$+$reverse shock model to both light curves as described in Section \ref{sub:GRB160623A}. }
  \label{fig:160623A_ami_sma}
\end{subfigure}
\caption{The X-ray (upper subplot) and radio (lower subplot) light curves for GRB 160623A. The fit results are presented in Table \ref{tab:GRB160623A_params}. }
\label{fig:160623A}
\end{figure}

GRB 160623A was detected by the Fermi GBM on 23 June 2016 at 05:00:34.23 UT \citep{2016GCN.19553....1V}. \textit{Swift}-XRT performed follow up observations detecting a power law decaying component until $\sim$5\,days post-burst. The event has a redshift of 0.367 \citep{2016GCN.19708....1M}. We plot the X-ray light curve in Figure \ref{fig:160623A_lc_Xray} (the first three data points are average values of a larger cluster of data points). We find that the XRT light curve follows a decay rate of $t^{-2.57\pm0.08}$. The \textit{Swift} Burst Analyser page for GRB 160623A quotes a photon index of $\Gamma = 1.8\pm0.2$ corresponding to a spectral index of $\beta_{\textrm{X}} = -0.8\pm0.2$.

Comparing the X-ray data to analytical models, the X-ray light curve for GRB 160623A is too steep for `regular' forward shock evolution. It is most consistent with jet break predictions \citep{1999ApJ...519L..17S} where $\alpha_{\rm{jet}} = -p$ so $p = 2.57\pm0.08$ for any emission above $\nu_{\rm{sa, m}}$. From the spectral index, we obtain $p = 2.6\pm0.4$, for $\beta = -0.8\pm0.2 = (1-\rm{p})/2$. So the X-ray emission is most consistent with a post-jet break scenario.

%is most consistent with optically thin forward shock emission (below the cooling break) in a stellar wind environment (i.e. $\rho \propto r^{-2}$), where we obtain $p = 2.85\pm0.09$ from the light curve and $p = 2.6\pm0.4$ from the spectral index ($\beta = -0.8\pm0.2 = (1-\rm{p})/2$). The X-ray light curve is too steep originate to be due to a homogeneous environment on the same branch of the synchrotron spectrum as the our spectral index and is too shallow to be in the regime $\nu_{\rm{c}} < \rm{X-ray}$. 

%If the spectral index $-0.8\pm0.2 = (1-\rm{p})/2$, then $p = 2.6\pm0.4$. From the power law decay of the X-ray light curve, we infer $p = 3.52\pm0.09$ and $p = 2.85\pm0.09$ in a homogeneous and stellar wind circumburst environments respectively. The stellar wind scenario is most of consistent with the value of $p$ from the spectral index. 

AMI--LA started observing the position of GRB 160623A 2.0\,days post-burst \citep{2016GCN.19609....1M} and extended to 902\,days post-burst. Figure \ref{fig:160623A_ami_sma} shows the AMI--LA light curve, which is best described by a broken power law plus a constant offset (i.e. Equation \ref{eq:brkn_pwl}). The results of the fitting are shown in Table \ref{tab:GRB160623A_params}. There is some scatter at late times which could be in part due to the low significances of the detections we obtain. It is likely that the 0.10$\pm$0.01\,mJy constant component to emission from the GRB's host galaxy, as was also seen in GRB 190829A \citep{2020MNRAS.496.3326R} or to some other unrelated radio source that is projected as being nearby on the sky (given AMI--LA's 0.5\,arcmin resolution this is a possibility).

\begin{table*}[t]
    \centering
    \begin{tabular}{cccccc}
        \hline
         & A\,(mJy) & $\alpha_1$ & $\alpha_2$ & $t_b$ (days) & B\,(mJy) \\
        \hline
        
        AMI--LA & 7.3$\pm$0.4 & 0.43$\pm$0.09 & -1.11$\pm$0.03 & 6.1$^{+0.4}_{-0.3}$\,days & 0.10$\pm$0.01\\
        SMA & 18.4$^{+0.6}_{-0.7}$ & -0.72 $\pm$ 0.04 & - & - & -\\
        X-ray & (0.22$\pm$0.02)$\times10^{-3}$ & -2.57$\pm0.08$ & - & - & -\\

        \hline

        %Radio (forward shock) & 1.2$^{+0.2}_{-0.1}$ & 0.0* & $-1.89$* & 34$^{+3}_{-16}$ & - \\
        %Radio (reverse shock) & 3.4$^{+0.3}_{-0.5}$ & 0.5* & $-2.9^{+0.9}_{-0.5}$ & 8.1$^{+1.1}_{-0.3}$ & 0.17$\pm$0.01 \\

        \hline
    \end{tabular}
    \caption{Results of our model fit to the AMI--LA, SMA and X-ray light curves of GRB 160623A as modelled by Equations \ref{eq: PL} and \ref{eq:brkn_pwl}. }
    \label{tab:GRB160623A_params}
\end{table*}

The rise component of the AMI--LA light curve $\alpha_{1} = 0.43\pm0.09$, is consistent with forward shock emission in a homogeneous environment ($\nu_{\rm{sa}} < \rm{AMI-LA} < \nu_{\rm{m,c}}$). The post-peak decay follows the evolution of $\nu_{\rm{m}}$ causing the light curve peak. The post peak decay is much shallower ($\alpha_{2} = -1.11\pm0.03$, $\nu_{\rm{sa,m}} < \rm{AMI-LA} < \nu_{\rm{c}}$) than the decay observed with \textit{Swift}-XRT ($\alpha = -2.57\pm0.08$, Figure \ref{fig:160623A_lc_Xray}). Given that the X-ray spectral index $\beta = -0.8\pm0.2$, and light curve places the X-ray emission on the same branch of the synchrotron spectrum as the radio emission ($\nu_{\rm{sa,m}} < \rm{AMI-LA, X-ray} < \nu_{\rm{c}}$), then the two decay rates should match.

The disagreement between the radio and X-ray data leads us to search for other scenario's to explain our afterglow data. In addition to the AMI--LA rise component being consistent with a forward shock ($<$10\,days in Figure \ref{fig:160623A_ami_sma}), it is also in agreement with reverse shock emission in the same spectral regime ($\nu_{\rm{sa}} < \rm{AMI-LA} < \nu_{\rm{m}}$). However, we now find a disagreement between the post peak decay and predictions of reverse shock light curves ($\nu_{\rm{sa, m}} < \rm{AMI-LA}$). The decay post peak is too slow to be from solely a reverse shock emission.

To help explain the disagreement between the radio and X-ray light curves, we use the published light curve from the SMA at 230\,GHz \citep{2020ApJ...891L..15C}. The light curve is plotted using red square data points in Figure \ref{fig:160623A_ami_sma} and follows a single power law decay with $\alpha = -0.72\pm0.04$ which is consistent with neither the X-ray or AMI--LA light curves. The SMA light curve is not explained by any predictions for forward shock counterparts in an wind environment. In short, the AMI--LA, SMA and X-ray light curves all show different decay rates over the same time period.

We calculate the AMI--LA-SMA (15.5-230\,GHz) spectral indices for the four sets of data points where we have contemporaneous (within 10\% of t\textsubscript{mid}-t\textsubscript{0}) data points in both bands. The two-point spectral index remains constant between $\beta_{\rm{5 days}} = 0.24\pm0.05$ and $\beta_{\rm{12 days}} = 0.12\pm0.08$ around the time of the AMI--LA peak, are close to the spectral regime $\nu_{\rm{sa}} < \rm{AMI-LA, SMA} < \nu_{\rm{m}}$, ($\beta = 1/3$). Given the spectral index measurements and the 15.5\,GHz light curve peak around the same time, it is most likely that the peak in the AMI--LA light curve is caused by $\nu_{\rm{sa}}$ such that both AMI--LA and the SMA sit on the same spectral branch post-peak. 

However, there is still substantial disagreement between the AMI--LA and SMA decay rates. To explain the disagreements, we propose that the observed radio and (less so) sub-millimeter emission is a superposition of both a forward and reverse shock component. Whilst not in perfect agreement with theory, the SMA counterpart can (for the most part) be described by a reverse shock component $\nu_{\rm{sa}} < \rm{SMA} < \nu_{\rm{m}}$, and our AMI--LA-SMA spectral index measurements show that the early time AMI--LA data is also from the same spectral component and therefore also from a reverse shock. We propose that the early time AMI--LA peak is cause by $\nu_{\rm{sa}}$ passing through the 15.5\,GHz band. As $\nu_{\rm{m}}$ from the reverse shock moves from below the SMA band at 230\,GHz through AMI--LA band, it steepens both light curves, giving way to forward shock emission. The later time AMI--LA emission becomes forward shock dominated. 

We attempt to quantify this explanation by performing a joint fit to the AMI--LA and SMA data where both a thick shell reverse and forward shock in a stellar wind environment (from the SMA light curve), are contributing to both datasets as well as the constant component at 15.5\,GHz. We only fit for the light curve normalization and break times, using both the X-ray decay rate for the forward shock decay and theoretical predictions for the rises and reverse shock decay (we assume $p = 2.5$). 

The fit results are shown in Figure \ref{fig:160623A_ami_sma}. At both frequencies, the forward and reverse shocks are contributing almost equally. The reverse shock dominates the earliest data, including creating the sharp peak we observe at 15.5\,GHz. After, the forward shock quickly takes over. In the AMI--LA data, the forward shock is more dominate at later times compared to the SMA light curve. Despite the reverse shock peak at $4.0\pm0.3$\,days, we are still not able to fully recreate the AMI--LA light curve peak. The model peak is earlier than the observed peak, which is most likely a result from enforcing the afterglow evolution to match the theoretical models exactly. We do not require any jet break scenario to explain the radio and sub-mm emission and so our scenario is still not consistent with the X-ray observations.

%First, we refit the AMI--LA light curve with two separate broken power laws, one including a constant offset, where one broken power law relates to the forward shock (which dominates at X-ray energies) and a second is due to a reverse shock (as seen in the SMA band). 

%In the refit, we fix three of the four $\alpha$ parameters, both of the rise components ($0.0$ and $0.5$) and one of the decays such that it matches the X-ray light curve ($\alpha_2 = -1.89$). With this set up, each peak is caused by $\nu_{m}$ moving through the 15.5\,GHz band. We find that combined, two broken power law fit reproduces the data well, as demonstrated by the small residuals in the lower panel of Figure \ref{fig:160623A_lc_Xray}. The purple and blue posteriors shown in Figure \ref{fig:160623A_ami_sma} represent the reverse and forward shock components respectively, with the gray posteriors being the sum of the two. We compare $\beta_{2} = -2.9^{+0.9}_{-0.5}$ from the reverse shock component to theoretical decay values and find that we have reasonable agreement \citep{2014MNRAS.444.3151V}. For a thin shell and thick shell reverse shock, we measure p = 3.3$^{+1.0}_{-0.6}$ and p = 1.88$^{+0.6}_{-0.3}$, respectively, both of which are in agreement with p values from other reverse shock studies \citep[e.g.][]{2018ApJ...859..134L, 2020MNRAS.496.3326R}. Therefore, we find that the AMI--LA light curve is well described by a combination of forward and reverse shock emission as well as contribution from the host galaxy.  

Our interpretation differs slightly from \citet{2020ApJ...891L..15C}, who used the sub-millimeter array (SMA), optical and X-ray observations to interpret the afterglow emission as coming from a combination of a structured (narrow $+$ wide) jet. It invokes an early narrow jet break in the X-rays, requires $p < 2$ and the SMA data to be dominated by the wider jet at later times. 

%\begin{figure}
%    \centering
%    \includegraphics[width=\linewidth]{GRB160623A_ami_lc_resid.pdf}
%    \caption{The AMI--LA light curve for GRB 160623A. Overlaid is a 100 different samples of the posterior distributions from 2 broken power law $+$ constant component model fit to the data. The blue posterior corresponds to forward shock component, also observed at X-ray energies. The purple posterior corresponds to a second shock component as well as the host galaxy emission.}
%    \label{fig:160623A_lc}
%\end{figure}

\subsection{GRB 210610B}

GRB 210610B was a long-duration GRB discovered by the \textit{Swift}-BAT telescope on the 10$^{\rm{th}}$ June 2021 at 19:51:27 UT \citep{2021GCN.30170....1P}. It was accompanied with a by a bright optical counterpart leading to a large follow up effort. Spectroscopic observations placed GRB 210610B at a redshift of 1.13 \citep{2021GCN.30182....1F}.

\begin{figure}[!ht]
\begin{subfigure}{\linewidth}
  \includegraphics[width=0.95\linewidth]{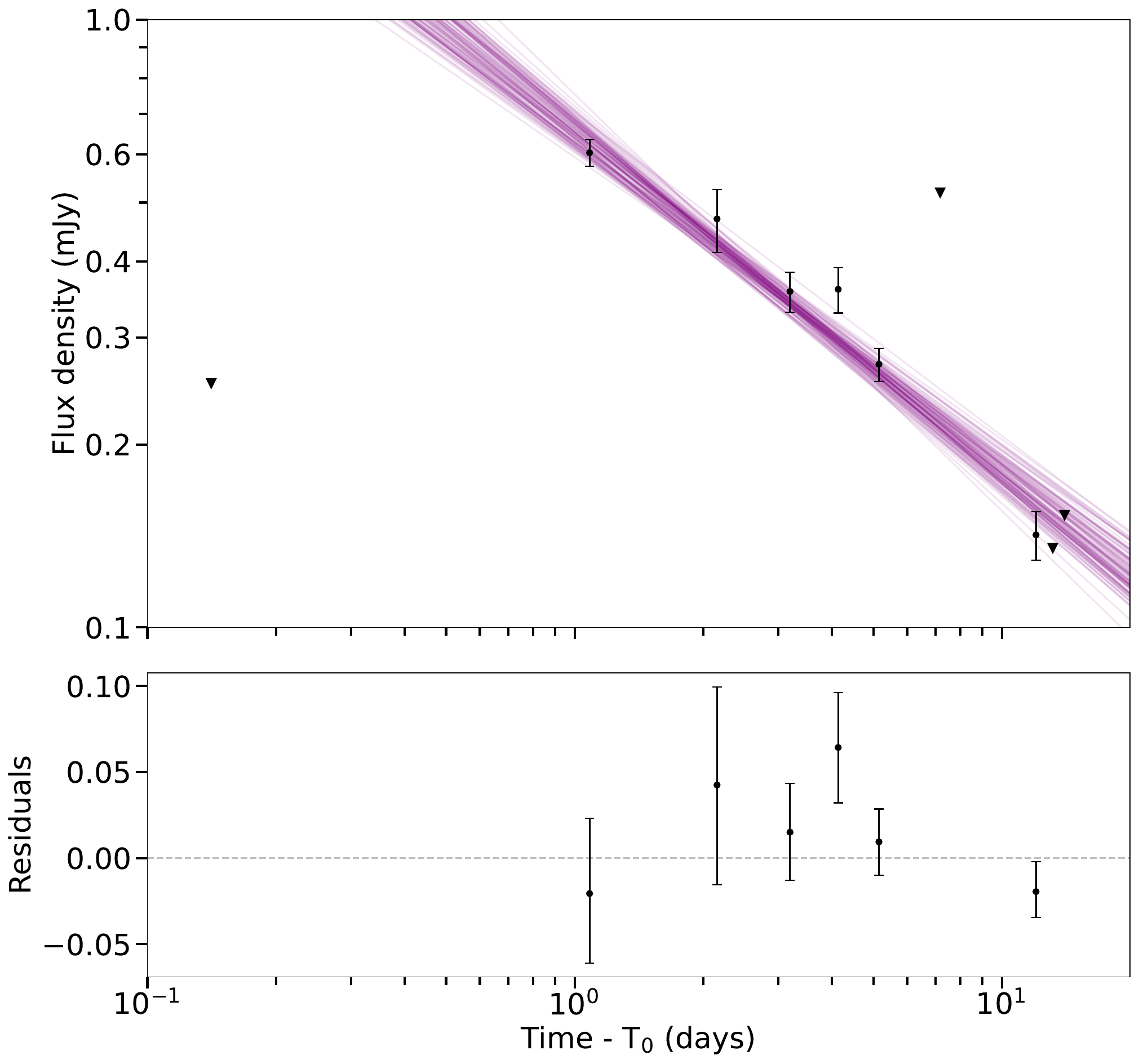}
  \caption{Light Curve of the AMI--LA data for GRB 210610B, with Equation \ref{eq: PL} fitted to the data, not including the early time upper limit. }
  \label{fig:Ami LC}
\end{subfigure}\hfill % <-- "\hfill"
~ % optional tilde b/t figures for readability.
  % this solution will not work w/ empty lines b/t subfigures
\begin{subfigure}{\linewidth}
  \includegraphics[width=0.9\linewidth]{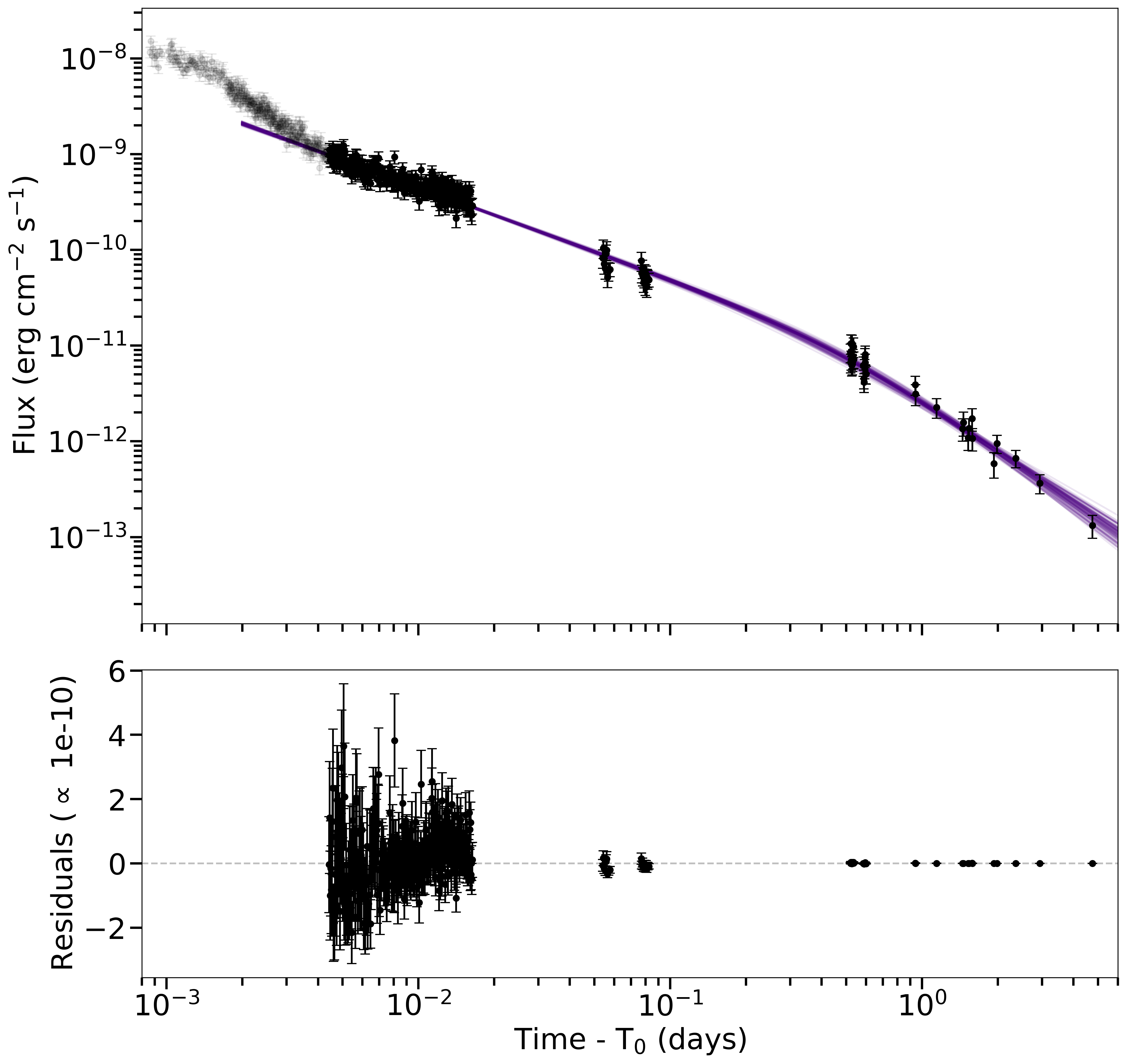}
  \caption{Light Curve of the Swift-XRT X-ray data for GRB 210610B, with Equation \ref{eq:brkn_pwl} fitted to the data.}
  \label{fig:Swift LC}
\end{subfigure}
\caption{The radio (upper subplot) and X-ray (lower subplot) light curves for GRB 210610B. The results of the model fitting are presented in Table \ref{tab:light curves best fit params}.}
\label{fig:210610B}
\end{figure}

\begin{figure}[hbt]
    \centering
    \includegraphics[width=\linewidth]{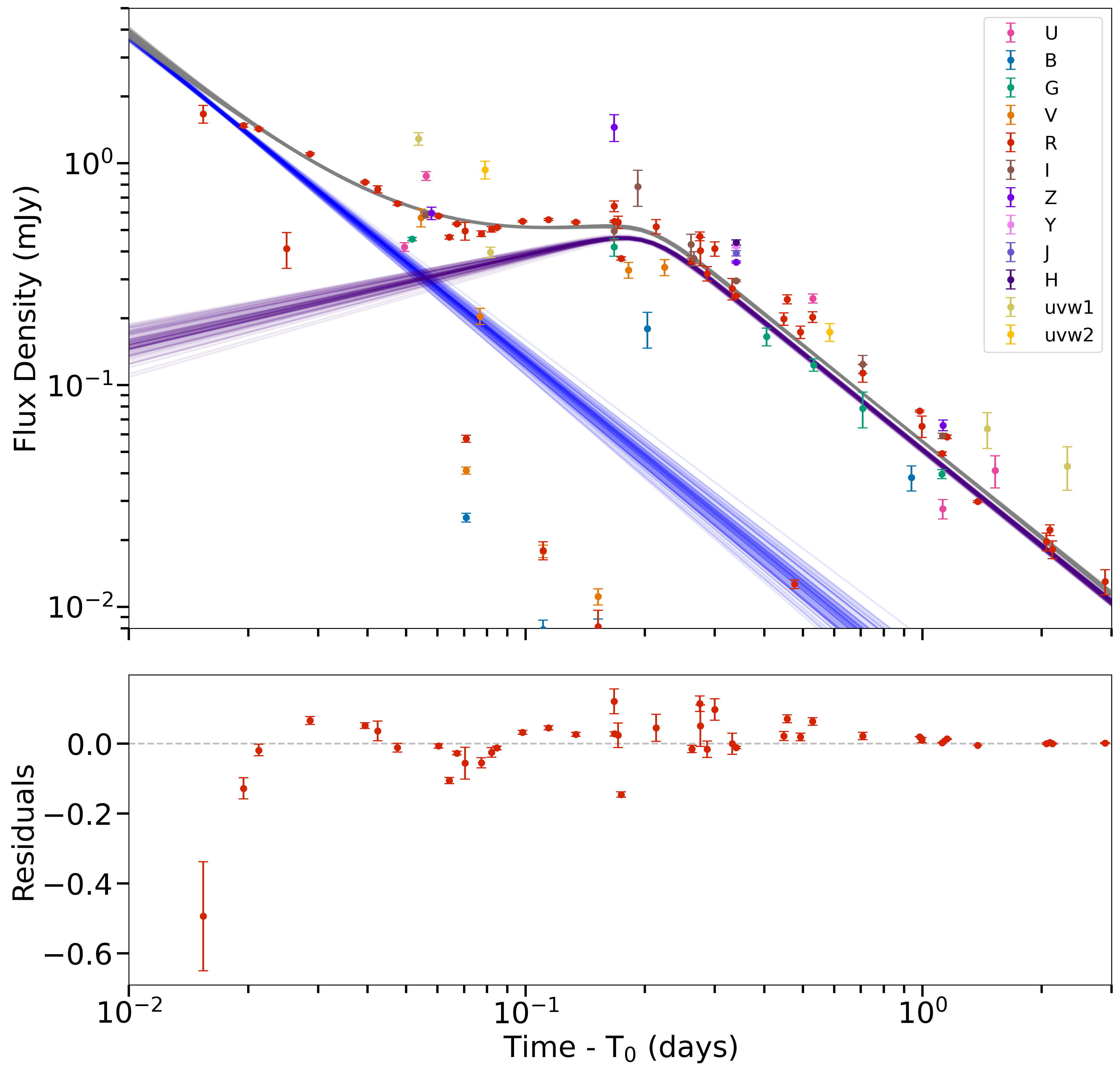}
    \caption{The optical data for GRB 210610B. We fit a single plus broken power law model to the  R-filter data. The blue and purple dashed lines represent, respectively, the smoothly broken power law and power law component of the fit, with the gray line as the combination of the two. The results of the model fit are presented in Table \ref{tab:light curves best fit params}.}%\citep{2024GCN.36556....1E,2024GCN.36559....1K,2024GCN.36561....1F,2024GCN.36562....1S,2024GCN.36573....1O,2024GCN.36574....1D,2024GCN.36575....1A,2024GCN.36576....1M,2024GCN.36577....1V,2024GCN.36579....1D,2024GCN.36582....1M,2024GCN.36585....1P,2024GCN.36589....1R,2024GCN.36592....1O,2024GCN.36597....1M,2024GCN.36599....1H,2024GCN.36601....1M,2024GCN.36613....1M,2024GCN.36654....1N,2024GCN.36655....1R,2024GCN.36656....1H,2024GCN.36734....1A,2024GCN.36947....1S}.}% The smoothly broken power law component has slopes $\alpha_1 = 0.40 \pm 0.04$ and $\alpha_2 = -1.44 \pm 0.01$, and break time $t_b = 0.200 \pm 0.002$ days. The power law component has a slope of $\alpha = -1.45 \pm 0.04$.}
    \label{fig:Red Filter LC}
\end{figure}

The full AMI--LA radio light curve is provided in Figure \ref{fig:Ami LC}. From 1 day post-burst (i.e. not including the early upper limit), it is best described by a single decaying power law: with $\alpha_{\rm{15.5GHz}} = -0.57 \pm 0.04$, which is too shallow to originate from a forward shock. Inputting the temporal index of the AMI light curve into the equations describing the decaying segments in the forward shock model spectrum yielded values of $p$ below 2, which is possible when invoking a high energy cut off. The temporal slope of the radio light curve is also consistent with the reverse shock model for a thick shell: falling within the expected range of -0.47 ($k = 0$) and -1.38 ($k = 2$) for the $\nu_{sa} < \nu < \nu_{m,c}$ segment of the reverse shock thick shell spectrum where the peak before one day post-burst is caused by $\nu_{\rm{sa}}$.% Comparison with the $F_\nu (\nu_{sa} < \nu_c, \nu_m < \nu_m, \nu_c < \nu)$ segment for k = 0, for which the temporal index is expected to be $\frac{94-73p}{96}$, yields a value of $p = 2.04 \pm 0.05$. This is within the expected range of $p \in [2, 3]$.  

The full \textit{Swift}-XRT light curve is given in Figure \ref{fig:Swift LC}. For GRB 210610B, we use the flux not flux density data because we found that substantially different results between the two, as a result of spectral evolution in the X-ray counterpart. The full light curve has four decaying power laws with different slopes. The very early-time data ($t < 0.004$ days) is not considered for the purpose of this study as it most likely comes from the prompt emission of the burst. The Swift-XRT X-ray light curve after $t = 0.004$ days is characterized by two decaying power laws, with a break happening at $t = 0.46 \pm 0.1$\,days. The best-fit parameters for the XRT light curve are reported in Table \ref{tab:light curves best fit params}. 

The X-ray emission of the pre-light curve break corresponds to segment $\nu_{sa,m} < \rm{X-ray} < \nu_c$ of the forward shock spectrum, where $p = 2.27 \pm 0.01$ and $k = 0$. The post-break decay is consistent with the $\nu_c < \rm{X-ray}$ where $p = 3.09 \pm 0.01$ and $k = 0$. Either side of the break, there is a drastic change in $p$, could (1) indicate that we are not fully correcting for other emission components at early times or (2) that the post-break decay is due to a jet break where $\alpha = -p$, where $p = 1.82\pm0.01$ or $\alpha = -3p/4$, where $p = 2.43\pm0.01$ depending on the jet model \citep{1999ApJ...519L..17S, 2013NewAR..57..141G}. We use the late time X-ray photon index $\Gamma = 1.9\pm0.1$ to break this degeneracy. The photon index corresponds to $p = 2.8\pm0.2$ ($\rm{X-ray} < \nu_{\rm{c}}$) or $p = 1.8\pm0.2$ ($\rm{X-ray} > \nu_{\rm{c}}$), the latter is consistent with a $\alpha = -p$ jet break, leading us to conclude that the break in the X-ray light curve at 0.04\,days is a jet break.

In addition to the radio and X-ray data, we have also compiled optical data from GCN notices and these are shown in Figure \ref{fig:Red Filter LC} \citep{2024GCN.36556....1E,2024GCN.36559....1K,2024GCN.36561....1F,2024GCN.36562....1S,2024GCN.36573....1O,2024GCN.36574....1D,2024GCN.36575....1A,2024GCN.36576....1M,2024GCN.36577....1V,2024GCN.36579....1D,2024GCN.36582....1M,2024GCN.36585....1P,2024GCN.36589....1R,2024GCN.36592....1O,2024GCN.36597....1M,2024GCN.36599....1H,2024GCN.36601....1M,2024GCN.36613....1M,2024GCN.36654....1N,2024GCN.36655....1R,2024GCN.36656....1H,2024GCN.36734....1A,2024GCN.36947....1S}. The R-band light curve is best described by two decaying power laws connected by a plateau segment from 0.07 to 0.2 days. We fit the R-band light curve using two components: a single and a broken power law. The results of the different slopes and break times are reported in Table \ref{tab:light curves best fit params}.% as $\alpha$, $\alpha_1$, $\alpha_2$ and $t_b$.

%The best-fit to the optical light curve was found using an adjusted version of Equation \ref{eq:brkn_pwl}, with added factors of $10^{-7}$ and $10^{-5}$ in front of $A$ and $B$, to avoid fit errors caused by very small numbers. The priors used were:  $0 < A (\rm{mJy}) < 1,000$, $-3 < \alpha < 0$, $0 < B\rm{mJy} < 1,000$, $0.01 < t_b < 1$, $ 0< \alpha_1 < 3$ and $-3 < \alpha_2 < 0$.  

\begin{table*}[t]
    \centering
    \begin{tabular}{ccccccc}
    \hline
         & A (mJy) & $\alpha$ & B (mJy) & $\alpha_1$ & $\alpha_2$ & $t_b$ (days) \\
        \hline
         AMI--LA & 0.67$\pm$0.04 & -0.58 $\pm$ 0.04 & - & - & - \\
         R-band & 45$^{+8}_{-7}$& -1.45 $\pm$ 0.04 & 44.4$\pm$ 0.4 & 0.40 $\pm$ 0.04 & -1.44 $\pm$ 0.01 & 0.200 $\pm$ 0.002\\
         X-ray & - & -0.95 $\pm$ 0.02 & - & -1.8 $\pm$ 0.1 & 0.5 $\pm$ 0.1 & -\\

         \hline
    \end{tabular}
    \caption{The parameters best describing the AMI--LA 15.5 GHz, R-band and X-ray 0.2 - 10 keV light curves for GRB 210610B, as modelled by Equations \ref{eq: PL} and \ref{eq:brkn_pwl}. We do not provide the normalisations for the X-ray fits because they were performed in flux space not flux density space. }
    \label{tab:light curves best fit params}
\end{table*}

%\begin{figure*}[htb]

%\begin{subfigure}{.475\linewidth}
%  \includegraphics[width=\linewidth]{GRB210610B_SED1.png}
%  \caption{SED 1: [0.94, 1.1] days}
%  \label{SED1 fit}
%\end{subfigure}\hfill % <-- "\hfill"
~ % optional tilde b/t figures for readability.
  % this solution will not work w/ empty lines b/t subfigures
%\begin{subfigure}{.475\linewidth}
%  \includegraphics[width=\linewidth]{GRB210610B_SED2.png}
%\caption{SED 2: [1.863, 2.277] days}
%  \label{SED2 fit}
%\end{subfigure}

%\medskip % create some *vertical* separation between the graphs
%\begin{subfigure}{.475\linewidth}
%  \includegraphics[width=\linewidth]{GRB210610B_SED3.png}
%  \caption{SED 3: [2.798, 3.419] days}
%  \label{SED3 fit}
%\end{subfigure}\hfill % <-- "\hfill"
% or just a comment and no tilde
%\begin{subfigure}{.475\linewidth}
%  \includegraphics[width=\linewidth]{GRB210610B_SED4.png}
%  \caption{SED 4: [4.563, 5.578] days}
%  \label{SED4 fit}
%\end{subfigure}
%\label{fig:sed_210610B}
%\caption{Plots of the four spectral energy distributions showing the possible range of values for the slopes of the segment of the spectrum each observing band is in. For both, p has a fixed value of 1.86 (+0.58, -0.54).}
%\end{figure*}

When considered independently, all three slopes obtained from the R-filter light curve are each consistent with the forward shock model. The first decaying component evolves as $F \propto t^{-1.45\pm0.04}$, indicating that this segment of the light curve could be in either the $\nu_m, \nu_{sa} < \rm{R-band} < \nu_c$ or $\nu_c < \rm{R-band})$ segment of the forward shock spectrum, regardless of the value of $k$. %Considering the scalings $\frac{3(1-p)}{4}$, $\frac{1-3p}{4}$ or $\frac{2-3p}{4}$ results in values of $p$ equal to $2.93 \pm 0.05$, $2.27 \pm 0.05$ or $2.60 \pm 0.05$. 

The rising component, which has a temporal index of $0.40 \pm 0.04$, is in agreement with the $\nu_{sa} < \rm{R-band} <\nu_m$ for a homogeneous environment. The second decaying component has a power-law index of $-1.44 \pm 0.01$ and is consistent with optically thin decaying forward shock model with p = 2.08 -- 2.25 for $k = 0$ and $k = 2$, respectively. %We do not consider $nu_{c} < \rm{R-band}$ because it would not reproduce the peak we see at 0.2\,days.

While the value of all three optical temporal slopes agrees with the forward shock model, the order in which they appear in the light curve does not. For the forward shock scenario we would require $\nu_{m}$ to travel both up and then down in frequency space through the optical observing band. 

One possible explanation for the presence of this first decaying segment is the detection of reverse shock emission in the optical band, with the reverse shock dominating at earlier times ($t < 0.1$ days) and the forward shock dominating at later times ($t > 0.1$ days). The temporal index of the initial R-band decay ($\alpha = -1.45 \pm 0.04$) is consistent with the expected values for forward shock and an optically thin ($\nu_{sa, m} < \rm{R-band} < \nu_{c}$), thick shell reverse shock in a homogeneous environment (to be consistent with the rest of the data) where $p = 2.74 \pm 0.04$.% is obtained when considering $\alpha$ to be $\frac{-3(5p-6)}{16}$, which is the 

At X-ray energies, we have suggested that the decay post-burst (0.46\,days) is consistent with a jet break. However this requires the optical decay to be the same on the same timescale given that jet breaks predict $\alpha = -p$ or $-3p/4$ for all $\nu > \nu_{\rm{obs}}$ i.e. either side of $\nu_{\rm{c}}$ \citep{1999ApJ...519L..17S, 2013NewAR..57..141G}. We do not find agreeing decay rates. However, the difference in the decay rates is $0.5\pm0.1$ consistent with $\nu_{\rm{c}}$ sitting between the optical and X-ray observing bands. Therefore, we conclude that there is no jet break observed in GRB 210610B. In order to rectify the AMI--LA observations with those at optical and X-ray energies, we infer that at least $\nu_{\rm{m}}$ (which causes the optical peak) sits between the optical and radio observing bands thus reducing the forward shock emission at low frequencies sufficiently such that the reverse shock is detectable in the radio band. The presence of reverse shock emission in our optical data helps lead to the conclusion that our radio data can also originate from a reverse shock component where $\nu_{\rm{m}}$ sits between the radio and optical observing bands. At X-ray energies, and for the later time optical emission, the forward shock dominates.

\subsection{GRB 210704A}

GRB 210704A was a long-duration GRB detected on the 4$^{\rm{th}}$ July 2021 at 19:33:45\,UT by the Fermi Gamma-ray Burst Monitor (GBM) and subsequently localized to arc-second scales by \textit{Swift}-XRT \citep{2021GCN.30380....1M, 2021GCN.30379....1D}. The Gran Telescopio Canarias (GTC) was used to measure a redshift of 2.34 \citep{2021GCN.30392....1D}. Subsequent multi-wavelength modeling of the afterglow found that the thermal explosion could not be explained by either supernovae or kilonovae models \citep{2023MNRAS.522.5204B}. It was bluer and faster-evolving than is usually observed for the supernovae associated with long GRB systems, leading to the conclusion that the explosion was more reminiscent of a luminous fast blue optical transient \citep{2026MNRAS.548ag555P}.

\begin{figure}
    \centering
    \includegraphics[width=\linewidth]{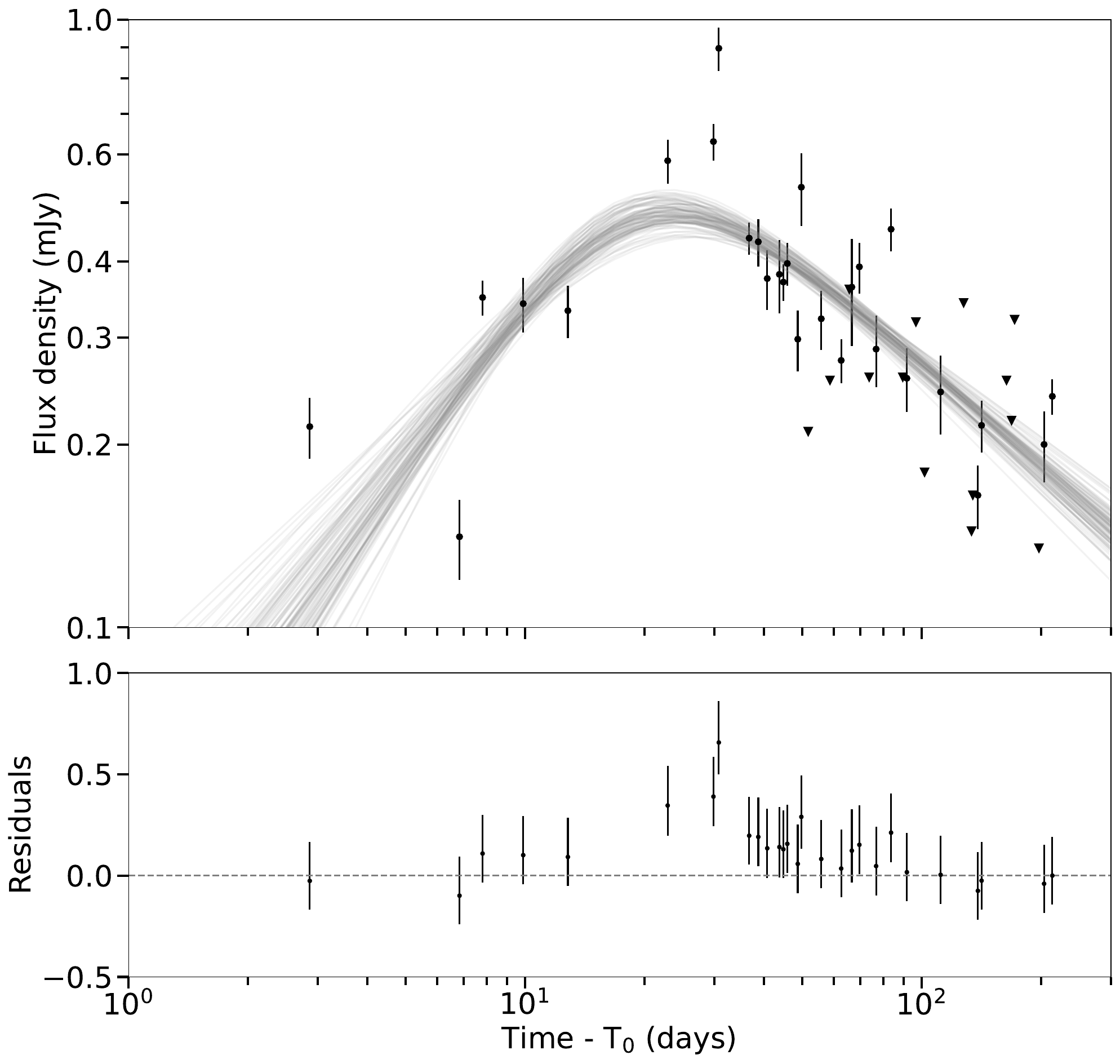}
    \caption{The AMI--LA light curve for GRB 210704A. There are large, rapid fluctuations in the light curve over the duration of our observing campaign which we attribute to scintillation. Overlaid is are 100 instances of the posterior distributions from a broken power law fit applied to the detections only. At all times there is large scatter about the posterior distribution.  }
    \label{fig:210704A_lc}
\end{figure}

Figure \ref{fig:210704A_lc} shows the light curve for GRB 210704A. The light curve can be described by a broken power law but the scatter is very large with fluctuations between deep upper limits and bright detections days of each other. We interpret the large, short timescale flux density changes to scintillation as a result of turbulence in the Milky Way’s interstellar medium. 

Along the line of sight to GRB 210704A, above 8.59\,GHz any observed scintillation is expected to be in the \textit{weak} regime \citep{2026ApJ..1002....3O}. In the weak scintillation regime, low level variability ($<<100\%$) is expected, as is seen in our observations. After de-trending our observations, i.e. removing the broken power law, we measure a fractional variability of (42$\pm$5)\% when including the upper limits \citep{2003MNRAS.345.1271V}. We can use the scintillation to place size limits on the afterglow component, at a redshift of 2.34 \citep{2021GCN.30392....1D}, the radio counterpart to GRB 210704A has to be smaller than $2.3\times10^{17}$\,cm for the duration of our observing campaign in order for the effects of scintillation to be observed. We do not perform any further modeling due to the large variability, the optical data is dominated by the thermal transient and there is limited X-ray data points, with little temporal overlap with the radio data.

%The variability observed is sufficiently large that we cannot confidently attribute to the radio emission to any given afterglow component.\textbf{not sure that I agree with that; given that there is other multi-wavelength data around, why did you decide not to do anything further with modeling of this source?}

\subsection{GRB 240529A}

\begin{table*}[ht]
\centering
\renewcommand{\arraystretch}{1.3}
\caption{Fitted parameters for GRB~240529A across different frequency bands. For the V and B-band data we populate the $\alpha_2$ column instead of $\alpha_1$ to make it easier for the reader to compare between the different results.}
\label{tb:LC_GRB240529A}
\begin{tabular}{lccccc}
\hline
 & $A\;(\mathrm{mJy})$ & $t_b\;(\mathrm{days})$ & $\alpha_1$ & $\alpha_2$ & $B\;(\mathrm{mJy})$ \\
\hline
Radio
    & $1.48\pm0.06$
    & $4.8\pm0.2$
    & $4.0^{+0.7}_{-0.8}$
    & $-0.95\pm0.04$
    & -- \\
R
    & $0.05538\pm0.004$
    & $0.116\pm0.005$
    & $3.2\pm0.2$
    & $-1.93\pm0.2$
    & $(3.9\pm0.2)\times10^{-2}$ \\
V& $0.146\pm0.007$
    & --
    & --
    & $-1.56\pm0.03$
    & -- \\
B
    & $0.075\pm0.008$
    & --
    & --
    & $-1.83\pm0.08$
    & -- \\
X-ray
    & $(10.6\pm0.6)\times10^{-3}$
    & $0.12\pm0.01$
    & $1.4\pm0.3$
    & $-1.91\pm0.03$
    & -- \\
\hline
\end{tabular}
\end{table*}
%\subsubsection{AMI--LA}

%AMI gcn \citep{2024GCN.36636....1R}

GRB 240529A was a long duration GRB detected by \textit{Swift}-BAT \citep{2024GCN.36556....1E} and reported to have a bright optical counterpart. It has a redshift of 2.035 \citep{2024GCN.36574....1D}.

\begin{figure}[!ht]
\begin{subfigure}{\linewidth}
  \includegraphics[width=0.95\linewidth]{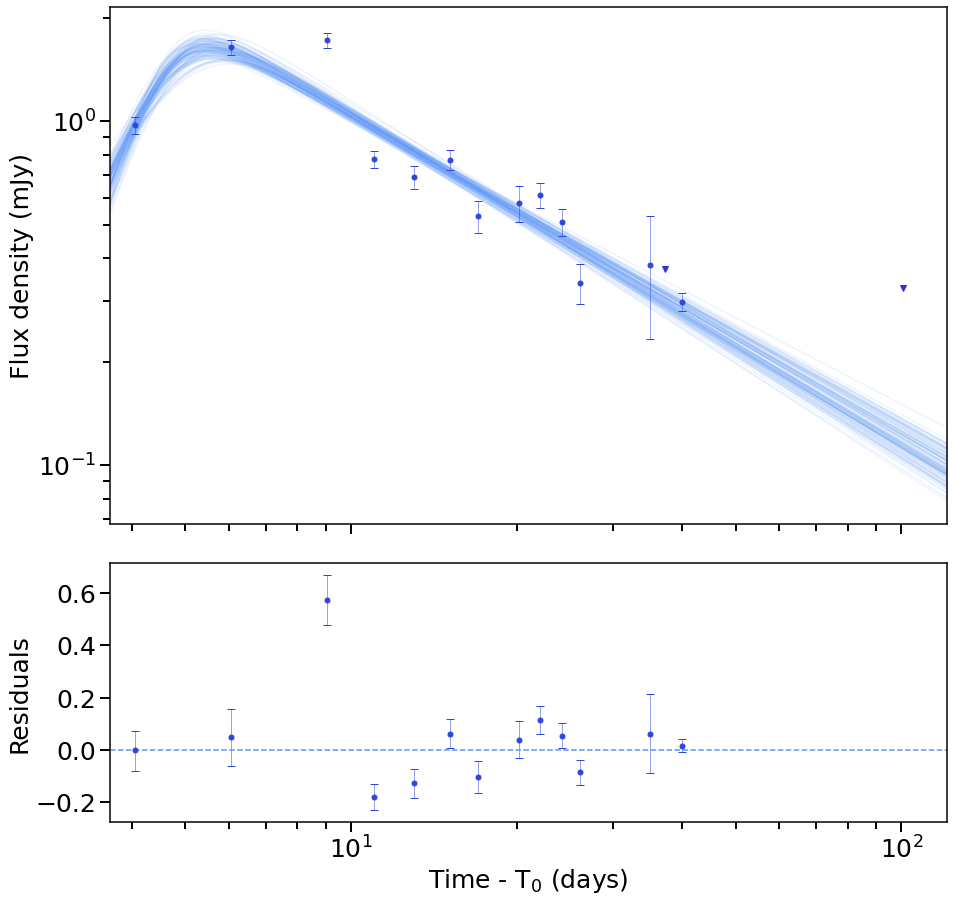}
  \caption{The AMI--LA light curve and model fit for GRB 240529A. }
  \label{fig: radio_GRB240529A}
\end{subfigure}\hfill % <-- "\hfill"
~ % optional tilde b/t figures for readability.
  % this solution will not work w/ empty lines b/t subfigures
\begin{subfigure}{\linewidth}
  \includegraphics[width=0.9\linewidth]{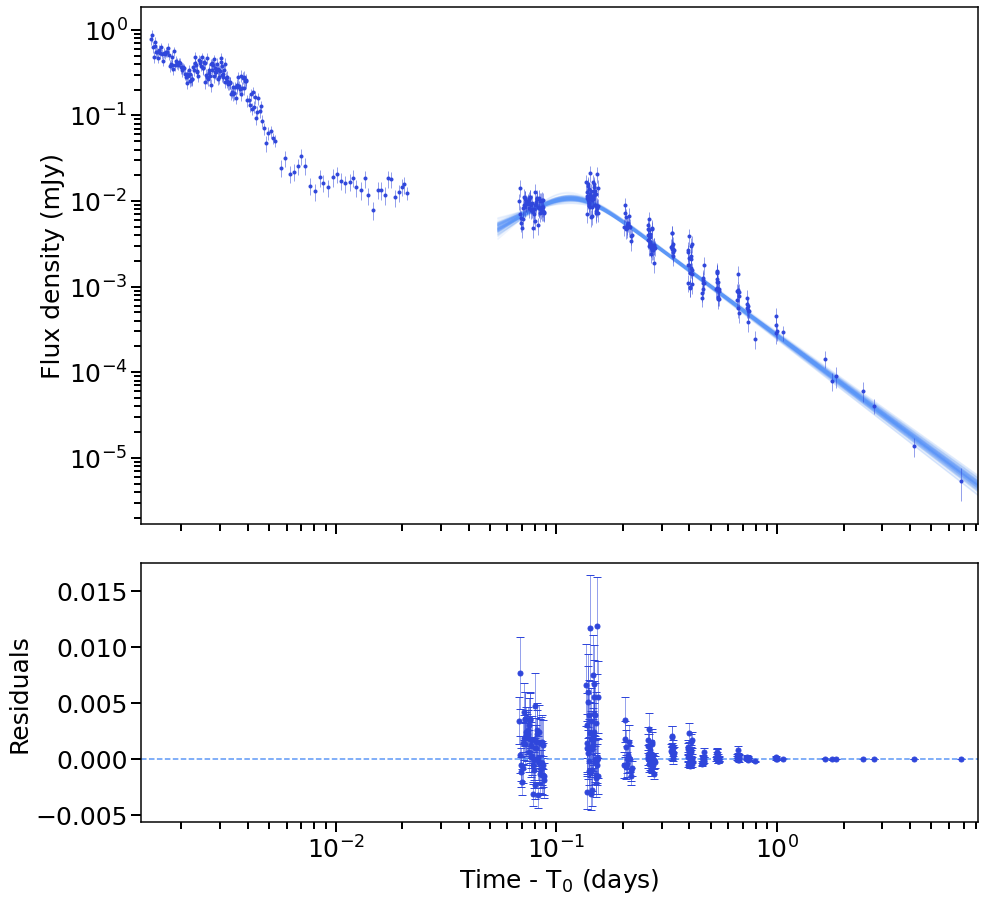}
  \caption{The Swift-XRT light curve for GRB 240529A. Overlaid is a 100 different samples of the posterior distributions from the broken power law model fit to the data. The first section of the data ($t<1$ hour) is believed to originate from the prompt emission and therefore is not part of the analysis.}
  \label{fig: xray_GRB240529A}
\end{subfigure}
\caption{The radio (upper subplot) and X-ray (lower subplot) light curves from GRB 240529A. The fitting results can be found in Table \ref{tb:LC_GRB240529A}.}
\label{fig:240529A}
\end{figure}

AMI--LA observed GRB 240529A in the radio range (15.5 GHz) from about $\sim4$ - 100\,days post-burst. The observations are shown in Figure \ref{fig: radio_GRB240529A}. We have fitted a smoothly broken power law (Equation \ref{eq:brkn_pwl}) to the data. Due to the lack of data from before the time break, we will exclude the first temporal index $\alpha_{1,\text{15.5GHz}}$ from our analysis. We found that $\alpha_{2, \text{15.5\,GHz}}=-0.95\pm0.04$ which is consistent with forward-shock emission in all both homogeneous and stellar wind environments as well as both below and above the cooling break ($\nu_{sa, m} < \rm{15.5\,GHz} < \nu_c$, $\nu_{sa, m, c} < \rm{15.5\,GHz}$). Given it is unlikely that the cooling break is below the AMI--LA observing band, we find that our AMI--LA decay is most consistent with $\nu_{sa, m} < \rm{15.5G\,Hz} < \nu_c$, in a homogeneous environment such that $p= 2.3\pm0.1$.

%The results are consistent with forward-shock emission in all three regimes, namely $\nu_m, \nu_c, \nu_{sa} < \nu = 15.5\,\mathrm{GHz}$. We therefore have our first estimate for $p$, the electron energy distribution index: $$\alpha_{2,\text{radio}}= -\frac{\left(3p-2\right)}{4}\implies p= 1.93\pm0.03.$$

\begin{figure}[ht]
    \centering
    \includegraphics[width=\linewidth]{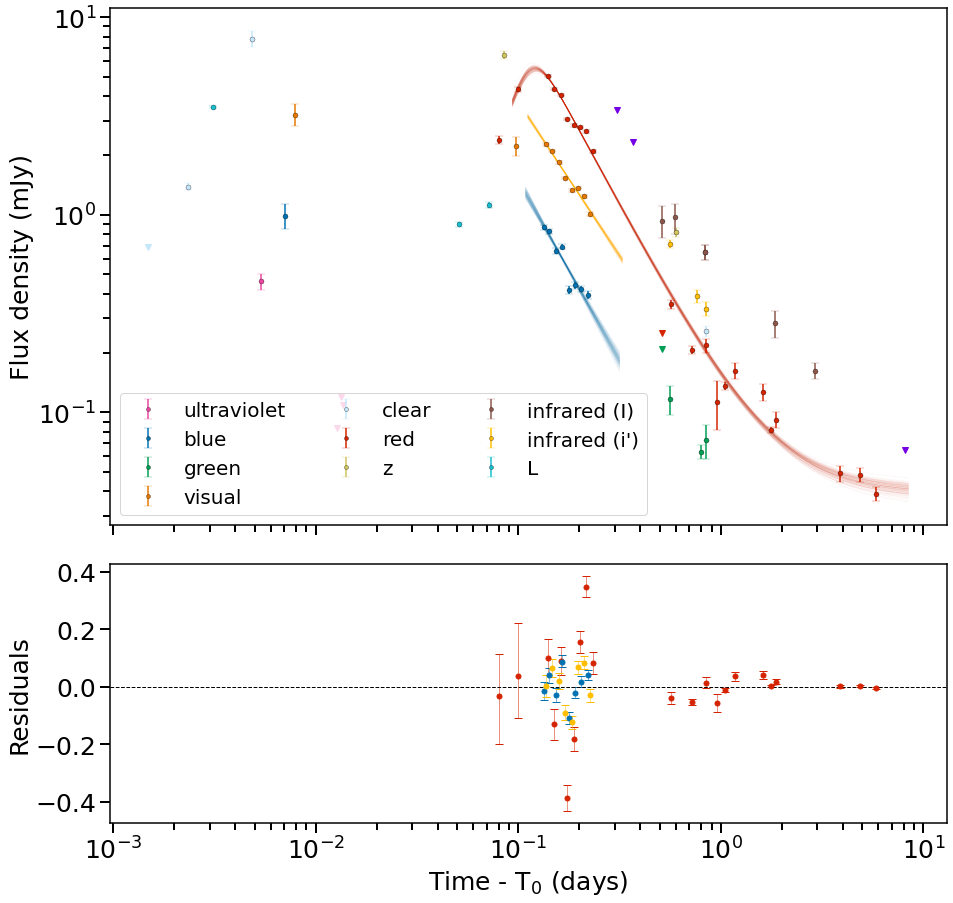}
    \caption{Optical light curve for GRB 240529A for all observing filters. For some frequencies, the data was fitted to smooth broken power laws or single power laws
    which are shown as a 100 overlaying different samples of the posterior distributions from the models fit to the data. %\citep{2024GCN.36556....1E,2024GCN.36559....1K,2024GCN.36561....1F,2024GCN.36562....1S,2024GCN.36573....1O,2024GCN.36574....1D,2024GCN.36575....1A,2024GCN.36576....1M,2024GCN.36577....1V,2024GCN.36579....1D,2024GCN.36582....1M,2024GCN.36585....1P,2024GCN.36589....1R,2024GCN.36592....1O,2024GCN.36597....1M,2024GCN.36599....1H,2024GCN.36601....1M,2024GCN.36613....1M,2024GCN.36654....1N,2024GCN.36655....1R,2024GCN.36656....1H,2024GCN.36734....1A,2024GCN.36947....1S}. 
    Fitting results can be found in Table \ref{tb:LC_GRB240529A}.}
    \label{fig: op_GRB240529A}
\end{figure}

As with GRB 210610B, we also consider optical data in our analysis. The optical observations, collected from GCNs \citep{2024GCN.36556....1E,2024GCN.36559....1K,2024GCN.36561....1F,2024GCN.36562....1S,2024GCN.36573....1O,2024GCN.36574....1D,2024GCN.36575....1A,2024GCN.36576....1M,2024GCN.36577....1V,2024GCN.36579....1D,2024GCN.36582....1M,2024GCN.36585....1P,2024GCN.36589....1R,2024GCN.36592....1O,2024GCN.36597....1M,2024GCN.36599....1H,2024GCN.36601....1M,2024GCN.36613....1M,2024GCN.36654....1N,2024GCN.36655....1R,2024GCN.36656....1H,2024GCN.36734....1A,2024GCN.36947....1S}, were first corrected for galactic extinction with E(B-V) = 0.332 (\cite{2024GCN.36562....1S}). For three specific observing bands with sufficient data, we fitted the data to either an broken power law plus constant component (fitted for $B$) for the R observing band (central wavelength $\lambda_R = 600$\,nm) or to a simple power law for the B and V observing bands ($\lambda_B= 415$\,nm, and $\lambda_V=520$\,nm, respectively). Early observations ($t < 0.125$\,days) in the blue and visual bands suggest that a time break occurs around $t = 0.125$\,days. However, these data were not included in the fit, as we were interested in the temporal indices after the break time, for which we fitted a simple power law. One data point in the red observing band ($t \approx 8$\,days) deviates from the general trend defined by the other observations. As it originates from a single telescope, it was excluded from the fitting procedure. The optical observations and fits for three filters (R, V, B) can be seen in Figure \ref{fig: op_GRB240529A}.

\textit{Swift}-XRT observed GRB 240529A from 0.001 -- 4\,days post burst (Figure \ref{fig: xray_GRB240529A}). The Swift-XRT observations reveal two plateau phases and three decay segments, including an early steep decay. We do not fit the data from less than 0.02\,days because we suspect it originates from the prompt emission and not the afterglow. Therefore, we fitted a broken power law fit with a single break (Equation \ref{eq:brkn_pwl}) to the later data. Similarly to the radio observations, we do not consider $\alpha_{1,\text{X-ray}}$ in our analysis because of the lack of data before the time break. If we follow regular forward shock emission, $\alpha_{2,\text{X-ray}} = -1.91\pm0.03$ gives us a value of $p$ close to $3$, either side of $\nu_{\rm{c}}$ for a stellar wind environment and $\rm{X-ray} > \nu_{\rm{c}}$ for a homogeneous environment ($\nu_{c} < \rm{X-ray}$, $p = 2.88\pm0.04$, independent of environment and $\nu_{c} > \rm{X-ray}$, $p = 3.21\pm0.04$. . %which is in disagreement with our results from the radio observations. 

The temporal index decays for the red and blue observing ranges ($-1.93\pm0.2$ and $-1.83\pm0.08$ respectively) are consistent with the X-ray decay rate. The temporal decay in the visual band ($-1.56 \pm 0.03$) is shallower than the others, which we conclude is likely explained by the limited amount of data. This interpretation is supported by the fact that, when we fitted a restricted subset of the R-band observations (corresponding to the size of the V observations), we obtained a shallower decay slope similar to that of the V band.

It is possible that the optical and X-ray data sit on the same branch of the synchrotron spectrum. We check this by calculating and comparing the optical optical-X-ray and X-ray spectral indices. Where we have multi-frequency optical data, we obtain spectral indices or around 5, indicating substantial extinction along the line of sight. The dust will referentially suppress high frequency optical emission. Therefore, we use R-band (i.e. lower frequency) data to calculate the optical-X-ray spectral index. We measure an optical-X-ray spectral index of $\beta_{\rm{opt-X-ray}} = -0.63\pm0.06$ after 0.12\,days by taking the average of each set of R-band and X-ray contemporaneous data points obtained within 5\% of the time post-burst of each other. The \textit{Swift} burst analyzer, reports a photon index of $\Gamma = 2.15^{+0.09}_{-0.05}$ corresponding to a $\beta_{\rm{X-ray}} = -1.15^{+0.09}_{-0.05}$. The difference between $\beta_{\rm{opt-X-ray}}$ and $\beta_{\rm{X-ray}}$ is $0.52^{+0.10}_{-0.08}$ in agreement with the expected different in spectral index either side of the cooling break. Therefore, the optical and X-ray emission must be on different spectral branches. To induce the same decay rate at both bands, we invoke a jet break at 0.12\,days post-burst.

%The flux decay and break time for the X-ray, red, blue and visual bands seem to all agree with one another, indicating that we have observed the jet break. 

%By combining all datasets and focusing on a small time interval in which the GRB was observed across many frequencies, we obtain the SED shown in Figure \ref{fig:SED GRB 240529A}. The X-ray observations have been averaged to a single point to avoid overlap.

%We now compare this spectral index with our observations. At around $t=97$ hours, we have observations in the radio band, X-ray band, and red band.  We use the $\beta_{\rm{X-ray}}$ from \textit{Swift}-XRT to constrain the high energy branch of the spectrum. Swift also provides the photon index ($\Gamma = 2.15^{+0.09}_{-0.05}$), which enables us to derive the corresponding spectral index $\beta$ for the X-ray data, shown on Figure \ref{fig:SED GRB 240529A} as an orange shaded region: $\beta = 1-\Gamma = -1.15^{+0.05}_{-0.09}.$ Extrapolation of the XRT spectrum to optical frequencies shows that the optical data point ($\nu \approx 10^6$ GHz) is too faint, \textbf{can it also be host galaxy extinction in the optical? if the X-ray and optical decays are similar, they are likely on the same spectral segment, unless we are indeed after the jet break; all of this needs much more careful treatment and wording} which indicates that there must be a frequency break between $\nu_{op} = 4.8\times 10^5$ GHz and $\nu_{Xray} = 2.4\times 10^9$GHz, corresponding to the cooling frequency $\nu_c$.  

%\textbf{REMOVE THIS PREVIOUS BIT? }

It is important to note that the decay rate is not shared by the radio observations as the temporal index $\alpha_{2,\text{radio}} = -0.95 \pm 0.04$ as observed at optical and X-ray bands. As previouly mentioned, some theories predict that jet breaks should be an achromatic process, however, observations have shown that radio counterparts often do not show the same break behavior \citep{2021ApJ...911...14K} and simulations have shown that close to the self-absorption break, the jet break can become chromatic and can delay the appearance of the break \citep{2011MNRAS.410.2016V}. Given we observed no jet break at 15.5\,GHz, this may be a possibility for GRB 240529A. % It has been noted multiple times in the GRB community that radio counterparts often do not show the same break behavior \citep{2021ApJ...911...14K}, although theory predicts that the break should be an achromatic process \textbf{this is not true if you are below or close to the self-absorption frequency, see van Eerten et al. 2011 (https://ui.adsabs.harvard.edu/abs/2011MNRAS.410.2016V/abstract)}, that is, it should occur across all wavelengths. 
Our findings here, reinforce the importance of late time radio observations in improving our understanding of how jet break signatures can varying as a function of observing frequency.

\subsection{GRB 240619A}
\label{sub:GRB240619A}

GRB 240619A was detected by Fermi-GBM at 03:43:31\,UT on 19$^{\rm{th}}$ June 2024 \citep{2024GCN.36694....1F}. The GOTO team found an optical counterpart enabling follow up \citep{2024GCN.36715....1G} and localization of the source to its host galaxy at z = 1.338 \citep{2024GCN.36813....1C}.

\begin{figure}[ht]
    \centering
    \includegraphics[width=\linewidth]{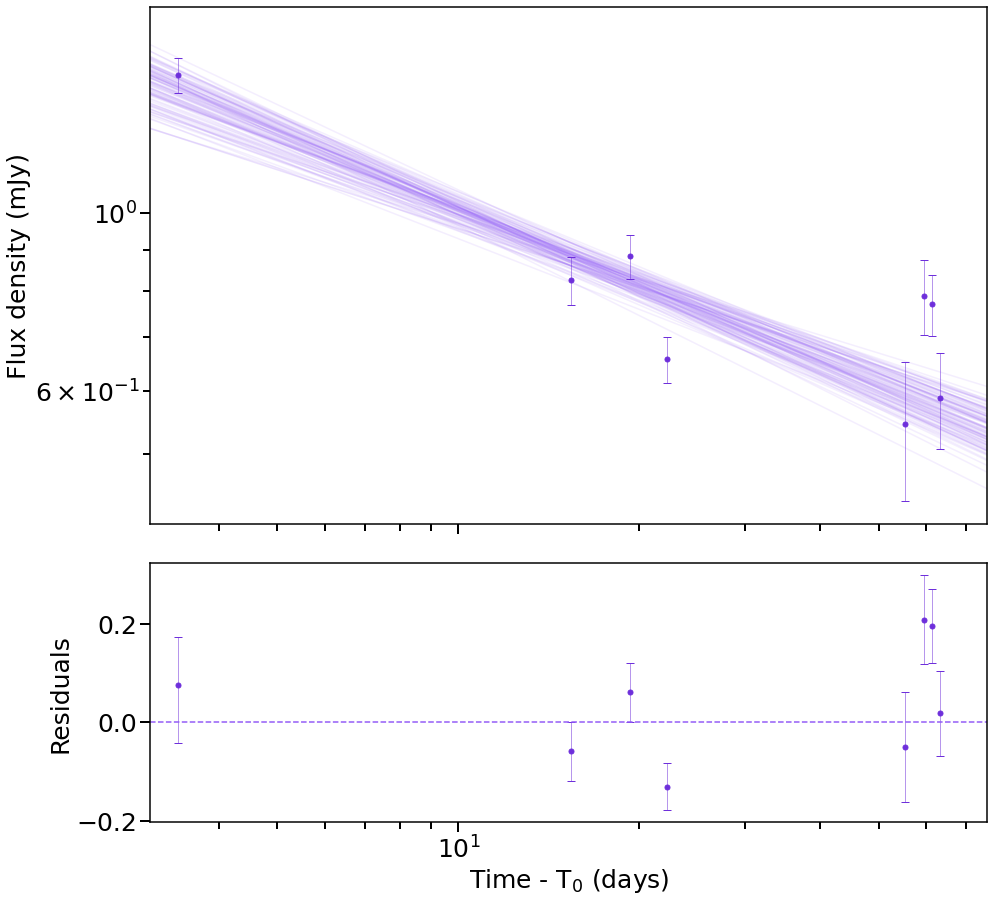}
    \caption{The AMI--LA light curve for GRB 240619A. Overlaid are 100 different samples of the posterior distributions from the broken power law model fit to the data. The model results are provided in the second paragraph of Section \ref{sub:GRB240619A}.}
    \label{LC AMI GRB 240619A}
\end{figure}

The radio light curve between 3 to 70\,days post-burst is shown in Figure \ref{LC AMI GRB 240619A}. It is best described by a decaying power law following $t^{-0.30\pm0.03}$, which is too slow to originate from a single forward shock component. As previously noted, GRB 240619A, was detected by Fermi, not \textit{Swift}-BAT, and while there were follow-up observations with \textit{Swift}-XRT observations, it is not possible to determine if the X-ray counterpart detected was varying. However, there was sufficient optical coverage to estimate the optical counterpart's decay rate and spectral slope \citep{2025MNRAS.544.1541K}. %We measure an optical decay rate of $\sim t^{-1}$ (we use `$\sim$' as this calculation used data from both broad and narrower observing bands with the roughly the same central frequency).

The optical decay rate is much steeper than what is observed at 15\,GHz \citep[$\alpha_{\rm{opt}} = -1$, ][]{2025MNRAS.544.1541K}. Comparison to theory shows that the optical decay is consistent with an optically thin decay (below the cooling break) in a homogeneous environment,  $\alpha = 3(1-p)/4 = -1$, therefore we obtain $p \approx 2.3$. However the radio decay rate is too shallow to originate from same emitting region because in a forward shock scenario, the light curve will only decay if the observing frequency is above $\nu_{\rm{sa}}$ and $\nu_{\rm m}$, which would mean it will decay at the same rate as at optical frequencies.

We also construct a 2-point spectral index with the optical and IR data from \citet{2025MNRAS.544.1541K} around 48\,days post-burst (when detections at multiple frequencies were obtained simultaneously) and measure a spectral index of $\beta \approx -0.75$. The optical spectral index and decay rate are both consistent with originating from the optically thin branch ($\nu_{\rm{sa, m}} < \rm{IR, optical} < \nu_{\rm{c}}$) of the synchrotron afterglow spectrum. Extrapolating the optical spectral index up to 15.5\,GHz drastically over predicts the AMI--LA flux density by nearly two orders of magnitude ($\sim100$\,mJy). Therefore there must be a spectral break between the radio and optical observing frequencies. If there is a spectral break, at least $\nu_{\rm{m}}$ and probably $\nu_{\rm{sa}}$ between the two bands so that 15.5\,GHz $< \nu_{\rm{sa, m}}$, the radio light curve should be increasing, which it is not. Therefore it is likely that the radio emission originates from a separate component to that shown by the optical light curve such as a reverse shock.

\subsection{GRB 251013C}

GRB 251013C was detected by both Fermi GBM and Space-based multi-band astronomical Variable Objects Monitor (SVOM) - ECLAIRS telescopes on 13$^{\rm{th}}$ October 2025 at 17:39:41\,UT \citep{2025GCN.42222....1R,2025GCN.42221....1F}. Given its low redshift of 0.572 \citep{2025GCN.42227....1M} it was subject to extensive multi-wavelength coverage. Here, we provide a brief overview of the radio counterpart as observed by AMI--LA. 

\begin{figure}[!ht]
\begin{subfigure}{\linewidth}
  \includegraphics[width=0.95\linewidth]{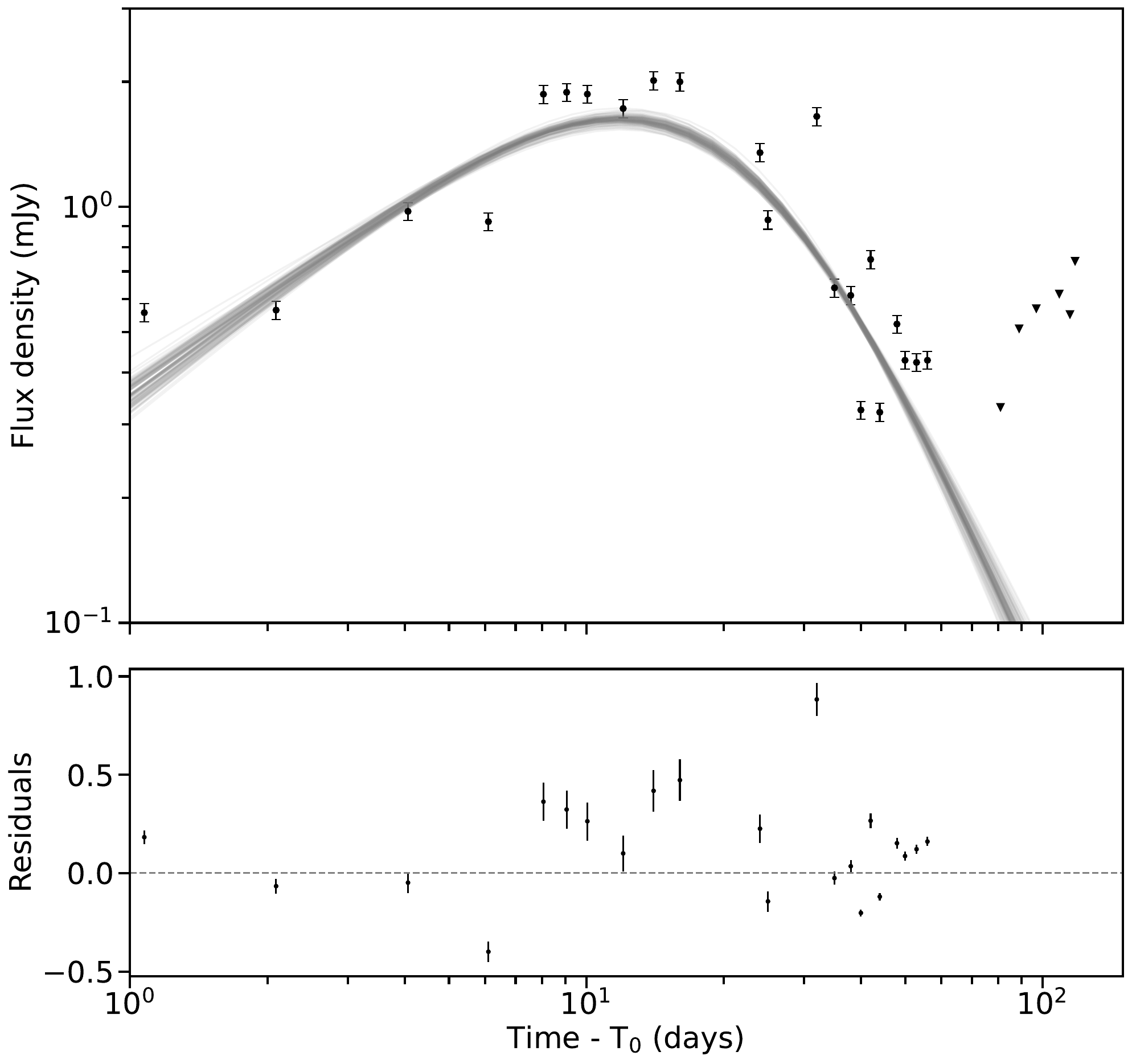}
  \caption{The AMI--LA radio light curve for GRB 251015C fitted with a broken power law fit. There is scatter in the residuals (on the bottom panel) that reduces slightly with time, after 30\,days post-burst, which we attribute to scintillation. }
  \label{fig:GRB251015C}
\end{subfigure}\hfill % <-- "\hfill"
~ % optional tilde b/t figures for readability.
  % this solution will not work w/ empty lines b/t subfigures
\begin{subfigure}{\linewidth}
  \includegraphics[width=0.98\linewidth]{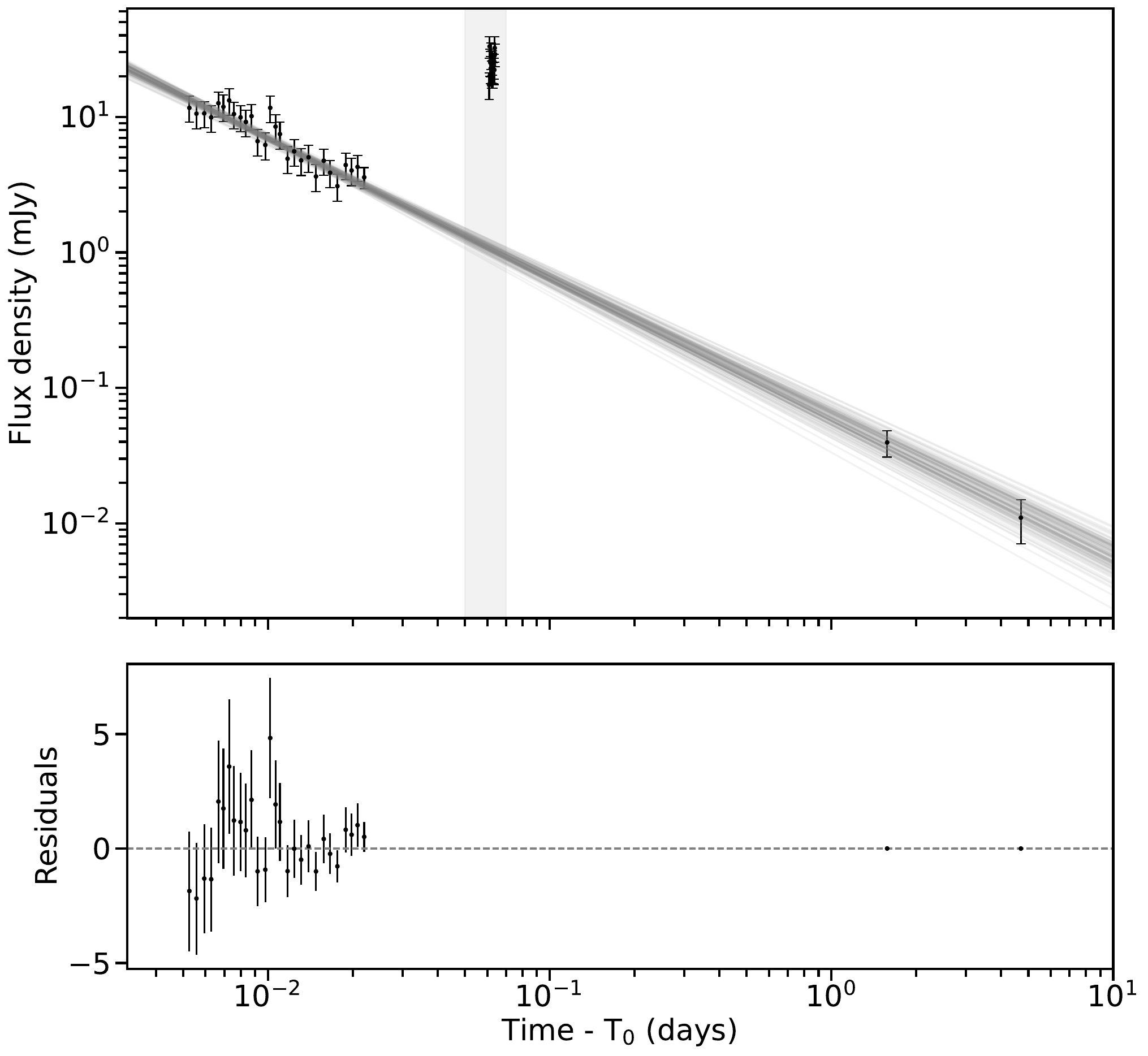}
  \caption{The \textit{Swift}-XRT light curve for GRB 251015C with a single power law fit to it. The vertical shaded region denotes a flare that we do not fit our single power law model to.}
  \label{fig: GRB251015C_Xray}
\end{subfigure}
\caption{The radio (upper subplot) and X-ray (lower subplot) light curves for GRB 251013C.}
\label{fig:251015C}
\end{figure}

The radio light curve for GRB 251013C (shown in Figure \ref{fig:GRB251015C}) is best described by a broken power law (see Table \ref{tab:params_251013C}). The top panel of Figure \ref{fig:GRB251015C} shows the model fit to the AMI--LA data. It is clear from the residuals in the bottom panel of Figure \ref{fig:GRB251015C} there is a some scatter about the fit to the data. The scatter is possibly the result of scintillation which means it is harder to trust the results of the broken power law fit. Not including the upper limits from $\sim$80\,days post-burst, we measure a variability amplitude of 29$\pm$1\% \citep{2003MNRAS.345.1271V}. For the line of sight to GRB 251013C, the transition frequency between a strong and weak scattering regime is 7.5\,GHz and so our observations at 15.5\,GHz are in the weak scattering regime \citep{2026ApJ..1002....3O}. The predicted modulation amplitude and variability timescale is 0.36 and 1.4\,days, respectively, roughly consistent with out observations given our observing cadence. Assuming weak scintillation, we place an upper limit of the radio emitting region source size of $3.6\times10^{16}$\,cm \citep[assuming a redshift of 0.572,][]{2025GCN.42227....1M}.

Tentatively, we interpret the post-peak decay as a jet break. We are able to do this given the steepness of the decay $t^{\sim2.6}$, when compared to GRB afterglow models. Post-jet break, a GRB afterglow evolves much faster than pre-jet break, the change in the decay slope is greater than we believe to be possible for scintillation. The sharp decay is most consistent with a jet break where theory predicts the decay rate of $t^{-p}$ where $p$ is the electron energy spectral index (the same as measured from the light curves and SEDs in other sections of this work \citep{1999ApJ...519L..17S}.

Comparison the X-ray counterpart, which was detected between 0.001 and 10 days, shows a much shallower power law decay. Figure \ref{fig: GRB251015C_Xray} shows the X-ray light curve for GRB 251013C. When fitting a power law to the data, we mask out the data between 0.05 and 0.07\,days (highlighted with the gray vertical strip) when there is an obvious flare in the data. We find that the X-ray counterpart decays following $t^{ -1.02^{+0.03}_{-0.04}}$. Our measured decay rate is much shallower than what was observed with AMI--LA, indicating that the jet break occurred somewhere between 10 and 20\,days post-burst, after the X-ray observations end. The X-ray decay is in the best agreement with a pre-jet break decay for optically thin synchrotron produced if the jet is expanding adiabatically into a homogeneous environment. Comparison of the light curve decay with theory, in this scenario, results in $p = 2.36^{+0.04}_{-0.05}$. The photon index reported by the \textit{Swift} Burst Analyser: $1.8\pm0.1$, is consistent with the aforementioned scenario (optically thin synchrotron radiation below the cooling break), corresponds to $p = 2.6\pm0.2$. It is also in agreement with the decay rate measured from the AMI--LA decay.

The radio and X-ray data for GRB 251013C can both be described by a single forward shock component propagating through a homogeneous environment. Furthermore, unlike GRB 240619A, here we find that the radio behaviour post-jet break follows the predictions well. 

\begin{table}[]
    \centering
    \begin{tabular}{ccccc}
    \hline
         & A (mJy) &$\alpha_{1}$ & $\alpha_{2}$ & $t_{\rm{b}}$(days) \\
    \hline
        Radio & 1.28$\pm$0.05 & 0.8$\pm$0.04 & $-2.6\pm0.1$ & 21$\pm$1\\
        X-ray & 0.061$\pm$0.009& -1.02$^{+0.03}_{-0.04}$& - & -\\
    \hline
    \end{tabular}
    \caption{The parameters from our single and broken power law fits to the AMI--LA radio and X-ray data sets for GRB 251013C, respectively.}
    \label{tab:params_251013C}
\end{table}

\section{Discussion}\label{sec:disc}

We have presented the second AMI--LA GRB afterglow catalog, containing observations spanning ten years of telescope operations. We also present individual analyses of six bursts (Section \ref{sub:individual}). We find that our detection rate (41/210 GRBs $=$ 20\% detection rate) is lower than has been found by others, \citet{2012ApJ...746..156C, 2018MNRAS.473.1512A}, who report detection rates of 48\% and 22--28\% (for an unspecified completeness threshold) or 15\% (for a completeness threshold of 0.2\,mJy), respectively. Our reduced detection rate is most likely originating from the large number of upper limits from our rapid response program.

There are many reasons why we do not detect radio emission in the first 24\,hours, the radio afterglow has not `turned on' yet, it is self-absorbed, it has not evolved into the AMI—LA observing band or is simply too low luminosity to be detected by our facility.

We compare our lack of detections in the first 24\,hours with observations from other observatories. Our findings in the first 24\,hours post-burst are consistent with those made with ATCA of GRBs 241113A and 240205B \citep{2024GCN.38287....1A, 2026arXiv260319047C}, which has been used to detect radio emission at 5/9\,GHz at just hours post-burst at 42 and $\sim100$\,$\mu$Jy, respectively. Extrapolating the two ATCA detections at 5/9\,GHz to the AMI--LA observing frequency at 15.5\,GHz using a self-absorbed spectral index \citep{2023NatAs...7..986B}, we obtain flux densities of 160 and 400\,$\mu$Jy, respectively. Only the latter would be marginally detected by AMI--LA in the first day post-burst. Therefore, whilst it is possible to detect radio emission in the first day post-burst, it seems that two hours on source is not sufficient for a facility like AMI--LA to detect the very early time counterparts.

The implementation of rapid response observations enables faster turn around of observations, as shown in figure 4 of \citet{2018MNRAS.473.1512A} (and Figure \ref{fig: Time dist Anderson}) where the distribution of observations for AMI--LA peaked at $<$1 day post-burst. Conversely, in \citet{2012ApJ...746..156C}, it was 1--2\,days due to the lack of rapid response modes. The turn around time for observations has since increased on the world's most sensitive facilities as a result of increased competition and the need for human intervention. 

Given that AMI--LA is not sufficiently sensitive to detect radio counterparts in the first hour post-burst, we recalculate our detection fraction considering only the bursts observed after 1 day and find we detect 37/162 bursts (23\%). Our adjusted detection fraction is consistent with those from \citet{2012ApJ...746..156C} and \citet{2018MNRAS.473.1512A}. 

Over the first few days post-burst, our detection fraction increases rapidly, and we obtain detailed light curves of GRB afterglows that sit above 0.2-0.3\,mJy. Figure 4 of \citet{2012ApJ...746..156C} shows that their flux density distribution peaks around 150\,$\mu$Jy between five and ten days post-burst. Considering the spectral evolution between the majority of the \citet{2012ApJ...746..156C} observations (at 8.46\,GHz) and the AMI--LA observing frequency (15.5\,GHz), a spectral index of $1/3$, we'd expect a peak 15.5\,GHz flux density of at most 200$\mu$Jy, which is only just detectable to AMI--LA. For detected events, AMI--LA can be used to produce high cadence light curves. Such detailed observations of bright transients provides opportunities to extract the most new physics and perform the most detailed modeling. Therefore, it is vital to build and maintain smaller, dynamically scheduled radio interferometers that can monitor interesting, bright transients and, when necessary, completely override current observations to trigger on a new target \citep{2023A&G....64.6.24F}.

Within our AMI--LA GRB sample, we have sufficient coverage to perform detailed analysis of 10 events \citepalias{2019MNRAS.486.2721B, 2020MNRAS.496.3326R, 2023NatAs...7..986B, 2024MNRAS.533.4435R}. Of those 10, at least six show evidence of reverse shock emission. We cannot fully attribute the inference of reverse shocks to the extent or density of our observations given that GRBs 240529A and 251013C which have comprehensive coverage but we find are best described by single emission components. We note that GRB 210407A shows evidence of scintillation, and so it is not possible for us to infer the presence of a reverse shock, therefore our total number of six reverse shocks maybe a lower limit. For us to infer reverse shock emission in the majority of our well-sampled afterglows, the relative peak flux densities of the forward and reverse shock components must be very similar which is dictated by the ratio Lorentz factors of the fireball ejecta at the time of launch and at the reverse shock crossing time \citep{1999ApJ...520..641S}. Usually, reverse shock components are inferred in datasets with comprehensive multi-wavelength radio data \citep[in addition to optical and X-ray coverage, e.g.][]{2018ApJ...859..134L}. Conversely, we are able to show that reverse shock components contribute to the total radio emission more often than not in a catalog with coverage at a single radio frequency.

With such variation of afterglow light curves and a prevalence of reverse shock emission we can also search for diagnostics of reverse shock emission in the radio data. For the 10 events in which we can measure, or place limits on, of the peak time post-burst, we find that peak of the afterglow varies by nearly two orders of magnitude in time. The earliest peak we can measure is GRB 221009A at $\sim0.3$\,days \citepalias{2023NatAs...7..986B} and the latest is GRB 251013C at $\sim21$\,days. Between these two limits, we find a relatively flat distribution in peak times in log-space. When considering the interpretations of our well-sampled afterglow light curves, we find no strong preference for either reverse or forward shocks dominating earlier times, for example the radio light curves for both GRB 190289A and 180720B were decaying from the first observation at $\sim1$\,day post-burst but were interpreted as reverse and forward shocks, respectively \citep{2020MNRAS.496.3326R}. We note that for the later time peaks, there is more ambiguity where we find attributing the light curve peak to a single shock component more difficult.

\begin{figure}
    \centering
    \includegraphics[width=\columnwidth]{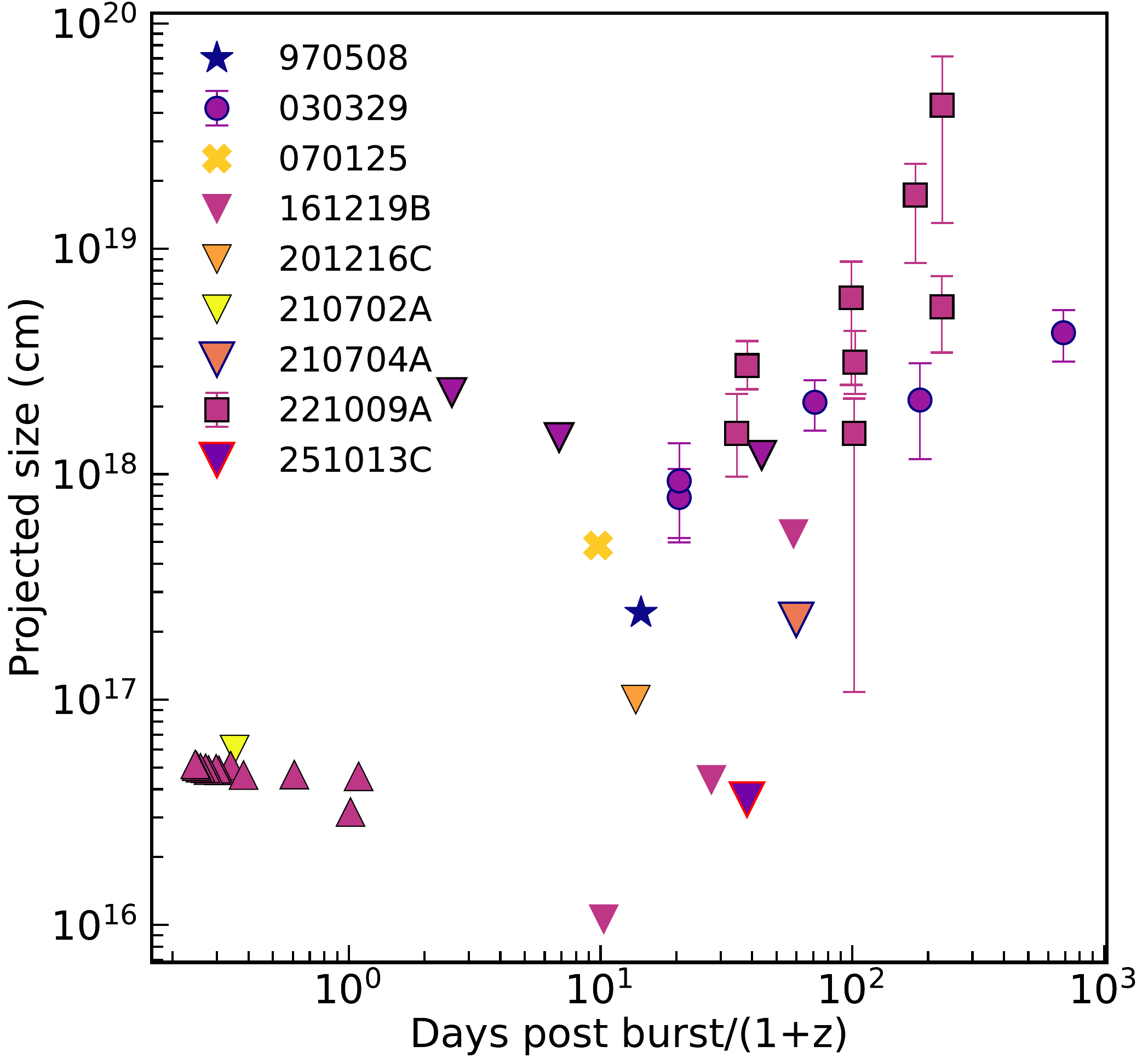}
    \caption{The source size limits for GRBs 210704A and 251013C via the detection of scintillation. Also shown are source size measurements and limits (upwards or downwards facing triangles) from the literature \citepalias{1997Natur.389..261F,2004ApJ...609L...1T,2005ApJ...622..986T,2007ApJ...664..411P,2008ApJ...683..924C,2012ApJ...759....4M,2019ApJ...870...67A,2022MNRAS.513.1895R,2023NatAs...7..986B,2023MNRAS.523.4992A,2024A&A...690A..74G}. Both of our limits are deeper that measurements made at the same timescale.}
    \label{fig:sizes}
\end{figure}

In addition to studying the emission processes and regions that dominate the afterglow emission, we have also been able to use these observations to help understand the jet hydrodynamics. Figure \ref{fig:sizes} shows the project source size vs rest frame time for nine different bursts from the literature including two of our afterglows that show scintillation (GRBs 210704A and 251013C). The size measurements and limits from the literature have been measured via scintillation (upper limits), equipartition analysis (lower limits) and VLBI (lower limits and detections of source size). Our limits for GRB 210704A and 251013C are the deepest limits on their respective timescales. In particular, the upper limit associated with GRB 251013C, is 1.5 orders of magnitude lower than the source size measurement associated with GRB 221009A on the same timescale. Interestingly, GRB 251013C shows `traditional' forward shock evolution including a jet break at 20\,days post-burst whereas GRB 221009A showed multiple emission components included a proposed wider outflow cocoon-type component which could explain the dramatic difference in source size measurements. We find the limit calculated for GRB 210704A also sits an order of magnitude lower than the size measurement for GRB 030329 on the same timescale. It appears that our observations of scintillation at 15.5\,GHz enable very deep constraints on the source size comparable to GRB 161219B \citep{2019ApJ...870...67A}. It is important to note that while we report limits, the value of these limits have large uncertainties associated with them stemmed from the scattering model used which relies on the Galactic electron distribution which in itself is still uncertain.

\begin{figure}
    \centering
    \includegraphics[width=\linewidth]{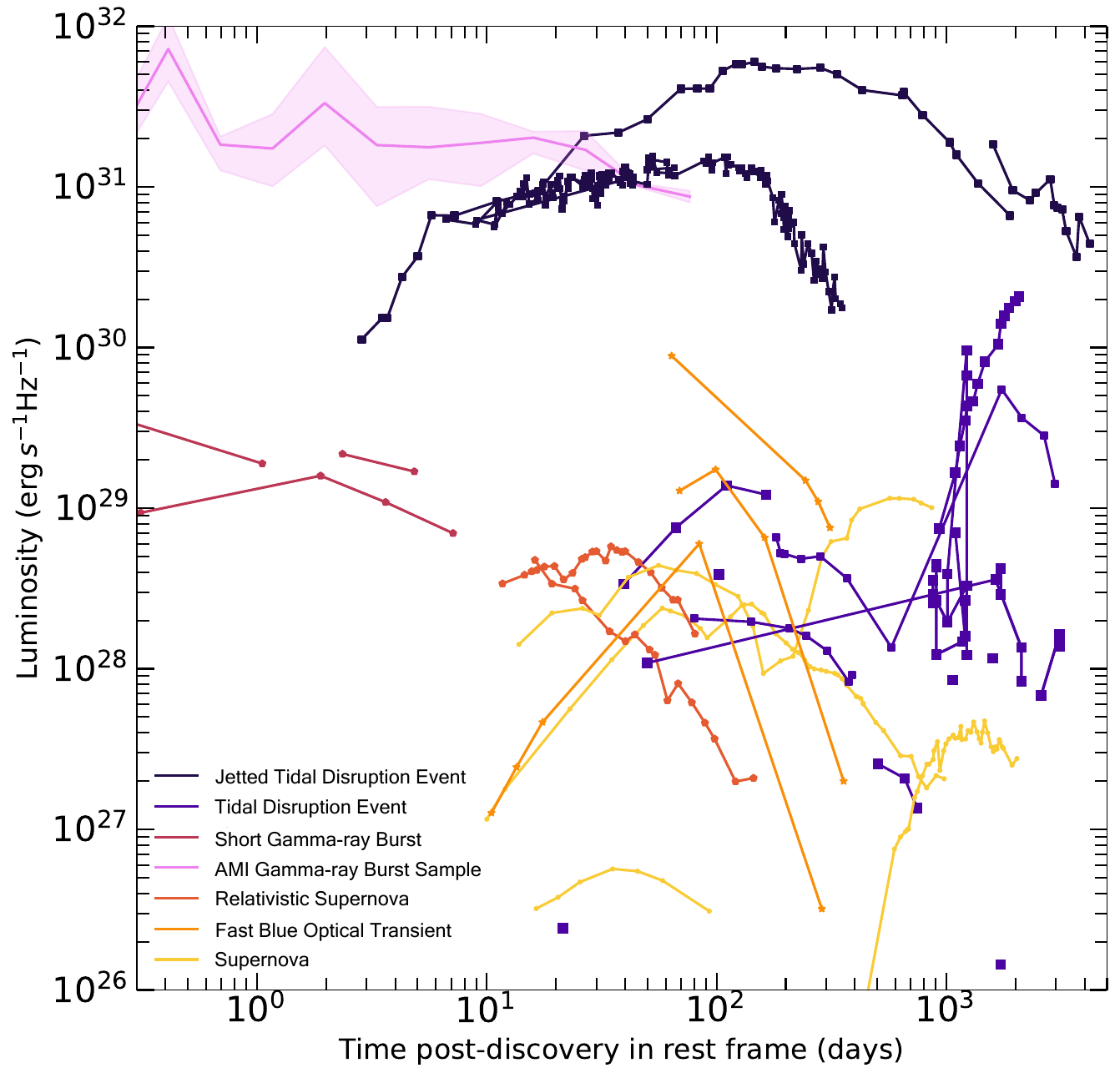}
    \caption{Average luminosity light curve of the AMI--LA GRB population sample compared to other transients. The pink light curve sitting in the top left corner is the average AMI-GRB-2 light curve. The other transients plotted are supernovae (SN), relative supernovae (Rel. SN), fast blue optical transients (FBOT), tidal disruption events (TDE), relative tidal disruption events (rel. TDE) and short gamma-ray bursts (short GRB). }
    \label{fig: Average Luminosity}
\end{figure}

The detected events within the AMI--LA catalog show large variance in their luminosity as well as peak timescale. Figures \ref{fig: Redshift Lightcurves} and \ref{fig: Average Luminosity} show our catalog in luminosity space. All of our radio-detected bursts have a redshift measurement. The range of luminosities at radio frequencies is larger than observed for the afterglow in other bands, spanning nearly five orders of magnitude (we emphasize that the afterglow because the prompt emission has an even broader luminosity distribution spanning at least seven orders of magnitude, \citet{2023ApJ...946L..31B}). \citet{2024MNRAS.533.4023D}'s optical afterglow sample spans about 3 order of magnitude in luminosity space at a given time post-burst. The \textit{Swift} afterglow sample shows that the luminosity range spans closer to two orders of magnitude at a given time \citep{2012MNRAS.425..506D}. The broader distributions of luminosities at radio frequencies at a given time step is most likely due to the broader range of behaviors seen at radio frequencies (as demonstrated in Section \ref{sub:individual}). We attribute the variance to (1) the radio properties are strongly dependent on the underlying jet parameters given both $\nu_{\rm{m}}$ and $\nu_{\rm{sa}}$ sit at radio frequencies and (2) we are likely observing multiple emitting regions which evolve on different timescales.%where we are likely observing a superposition of multiple emitting regions therefore afterglow can be brighter than if there was just a forward shock component. The larger variance at a given time step is also due to the radio properties strong dependence on the underlying jet parameters, given that both $\nu_{m}$ and $\nu_{sa}$ sitting at radio frequencies. 

The average light curve of our GRB population is plotted along with the light curve of other astrophysical transients in Figure \ref{fig: Average Luminosity}. The AMI--LA population sample sits in its own region of the transient phase space, the top left hand corner of Figure \ref{fig: Average Luminosity} with a luminosity between $10^{32}$ and $10^{31}$ erg s$^{-1}$ Hz$^{-1}$, the average light curve is rather flat because it averages over a broader range of behaviors including a range of peak times decay rates.%for a time post discovery fading from $0.3$ to $100$ days. %The light curve appears to be decaying over the time range post-discovery of our population, with a sharper decay beginning at $t \approx 15$ days. 

The only other synchrotron transient that sits at a similar apparent luminosity range are jetted or relativistic Tidal Disruption Events \citep[stars that are ripped apart by super massive black holes and launch relativistic jets that emit luminous radio emission, ][]{2018ApJ...854...86E, 2025ApJ...992..146R}. However, unlike the radio afterglows of GRBs, relativistic TDEs have rise timescales of $\sim$100\,days, (for GRBs it is up to 20\,days). As such it is possible to differentiate between jetted TDEs and GRBs based on their radio light curves. Whilst we do not expect a given GRB radio light curve to follow the path of the average light curve shown in Figure \ref{fig: Average Luminosity}, we might expect a given candidate GRB's light curve to intersect with this region of the luminosity-timescale parameter space. Such discriminators are important in the era of radio surveys that are enabling the discovery of new transients at radio frequencies as opposed to radio telescopes being a facility for follow up observations. %\textcolor{red}{citation}. \textbf{WHat does the average AMI light curve say?}

\section{Conclusions}\label{sec:conc}

In this paper, we have presented observations of 210 GRBs using the AMI--LA telescope spanning just around a decade. Of the data published, ten events have a sufficient number of detections that they can be analyzed in detail. Six of those ten events have not be published elsewhere. We have presented analysis of GRBs 150623A, 210610B, 210704A, 240529A, 240619A and 251013C, using their radio, X-ray and, in places, optical data to interpret the afterglow emission in the context of the fireball model. We find a broad range of behaviors, six of the ten events: GRBs 160623A, 171010A, 190829A, 210610B, 221009A and 240619A all deviate from a single emitting region model. Across the population it is clear that the radio counterparts to long GRBs, more often than not, require consideration of both forward \textit{and} reverse shock components. For those events where we detect reverse shock emission, we are finding that the reverse shock emission is contributing significantly for $\gtrsim$10 days. Therefore, we predict that as more sensitive radio facilities come online that are able to detect fainter afterglow components, it would become a necessity to automatically consider reverse shock emission in any afterglow modelling attempt.

We also find, via GRBs 240529A and 251013C, varying behavior in the radio afterglow light curves when jet breaks are observed at higher frequencies. It is possible that the variation can be attributed solely to the proximity of the AMI--LA observing band to the synchrotron self-absorption frequency \citep{2011MNRAS.410.2016V} however in order to determine this we would require multi-frequency observations. Therefore, such a study is left to future work. 

%In addition to considering individual bursts, we also examine our radio-observed GRB population as a whole. We find that the flux density distribution of our detections are flat down to the sensitivity limit of AMI--LA, indicating that our observations are sensitivity limited. The catalog considers both manually scheduled observations of new bursts as well as automated rapid response scheduling using ALARRM. We find that, whilst AMI--LA is able to get on source as soon as the trigger is received in some cases, the facility is not sufficiently sensitive enough to detect afterglows are such early times but we can place limits on northern hemisphere GRBs on timescales of a few hours post-burst. With small facilities such as AMI--LA is it possible to override observations to get on source of a GRB afterglow within minutes and continue to observe every day to track the full path that the afterglow takes. 

The second AMI--LA GRB afterglow catalog presented here has demonstrated that radio counterparts are diverse and broad ranging in their properties. Their variance is far greater than observed at any other frequency band. We have used the catalog to motivate the need for reverse$+$forward shock modeling of future GRB afterglows as a standard practice. We have shown that the high cadence light curves made using AMI--LA for GRBs has enabled science that would not have been otherwise possible. We advocate for continued funding of smaller radio facilities that can be dedicated to transient monitoring. With interferometers dedicated to transient science, we will continue to see our understanding of GRBs improve and be extrapolated to other physical systems. 

%, over an order of magnitude more than is observed at optical and X-ray energies. We hypothesise that this is also an effect of the contribution of multiple emitting components contributing to the radio counterparts. With an additional shock component there are more free parameters and therefore a larger range of parameter space in which the radio emission can vary. \textbf{Nope... the first thing to consider is the more complex spectral behavior at radio frequencies, before you start talking about multiple components...}

%% Please use the acknowledgment and contribution environments. This will 
%% be anonomyized when the "anonymous" style option is used. 
\begin{acknowledgments}

LR acknowledges funding from the Trottier Space Institute Fellowship, the Natural Sciences and Engineering Research Council of Canada (NSERC) Arthur B. McDonald Fellowship and Discovery Grant programs, the Canada Research Chairs (CRC) program, the Fondes de Recherche Nature et Technologies (FRQNT), the Centre de recherche en astrophysique du Québec (un regroupement stratégique du FRQNT), and the AstroFlash research group. The AstroFlash research group at McGill University, University of Amsterdam, ASTRON, and JIVE is supported by: a Canada Excellence Research Chair in Transient Astrophysics (CERC-2022-00009); an Advanced Grant from the European Research Council (ERC) under the European Union’s Horizon 2020 research and innovation programme (`EuroFlash’; Grant agreement No. 101098079); an NWO-Vici grant (`AstroFlash’; VI.C.192.045); an NSERC Discovery Grant (RGPIN-2025-06681); an ERC Starting Grant (`EnviroFlash’; Grant agreement No. 101223057); and an NWO-Veni grant (VI.Veni.222.295).

RF thanks UKRI, the ERC and The Hintze Family Charitable Foundation for support.

AKH thanks UKRI for support.

DH acknowledges support from the Natural Sciences and Engineering Research Council of Canada (NSERC) Discovery Grant and Arthur B. McDonald Fellowship programs, the Canada Research Chairs program, the Fonds de recherche du Québec via the Center for Research in Astrophysics of Québec (AstroQuébec; https://doi.org/10.69777/377645), and the Trottier Space Institute at McGill. 

AH is grateful for the support by the Israel Science Foundation (ISF grant 1679/23) and by the the United States-Israel Binational Science Foundation (BSF grant 2020203). 

KPM acknowledges the Anuandhan National Research Foundation Prime Minister’s Early Career Research Grant ANRF/ECRG/2024/005949/PMS.

We thank the Mullard Radio Astronomy Observatory staﬀ for scheduling and carrying out the AMI-LA observations. The AMI telescope is supported by the European Research Council under grant ERC-2012-StG-307215 LODESTONE, the
UK Science and Technology Facilities Council, and the University of Cambridge.

\end{acknowledgments}

%% To help institutions obtain information on the effectiveness of their 
%% telescopes the AAS Journals has created a group of keywords for telescope 
%% facilities.
%
%% Following the acknowledgments section, use the following syntax and the
%% \facility{} or \facilities{} macros to list the keywords of facilities used 
%% in the research for the paper.  Each keyword is check against the master 
%% list during copy editing.  Individual instruments can be provided in 
%% parentheses, after the keyword, but they are not verified.
\facilities{AMI--LA, the Neil Gehrels \textit{Swift} Observatory X-ray Telescope}

%% Similar to \facility{}, there is the optional \software command to allow 
%% authors a place to specify which programs were used during the creation of 
%% the manuscript. Authors should list each code and include either a
%% citation or url to the code inside ()s when available.
\software{\textsc{casa}, Python}

\bibliography{sample701}{}
\bibliographystyle{aasjournalv7}

%% Appendix material should be preceded with a single \appendix command.
%% There should be a \section command for each appendix. Mark appendix
%% subsections with the same markup you use in the main body of the paper.
%%
%% Each Appendix (indicated with \section) will be lettered A, B, C, etc.
%% The equation counter will reset when it encounters the \appendix
%% command and will number appendix equations (A1), (A2), etc. The
%% Figure and Table counter will not reset.

\appendix
\section{Analysis including the full AMI--LA catalogue} 
\label{sec:wholecatalogue}

Here, we reproduce Figures \ref{fig:full_cat}, \ref{fig: Flux Density Distribution}, \ref{fig:Time Since Burst Distribution}, \ref{fig:  Redshift Lightcurves} and \ref{fig: Redshift Distribution} including the first AMI--LA GRB catalog \citep{2018MNRAS.473.1512A}. In total both catalogs include 343 GRBs (totaling 1869 observations), 100 of which have redshifts.

\begin{figure}[h]
    \centering
    \includegraphics[width=\linewidth]{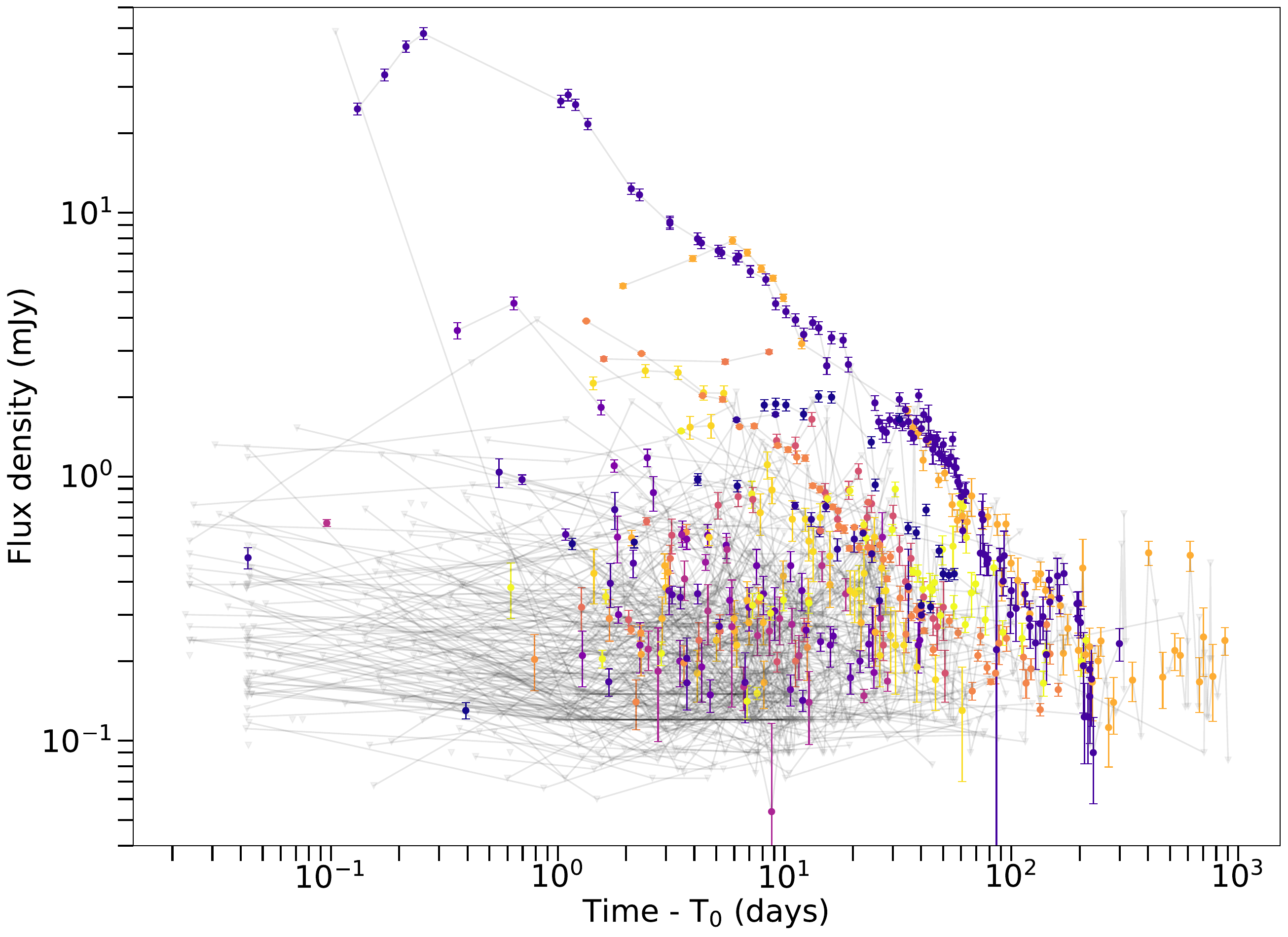}
    \caption{A full compilation of all observations of all GRBs from AMI-GRB-1 and AMI-GRB-2 \citep{2018MNRAS.473.1512A}.}
    \label{fig:total_sample}
\end{figure}

\begin{figure}
    \centering
    \includegraphics[width=\linewidth]{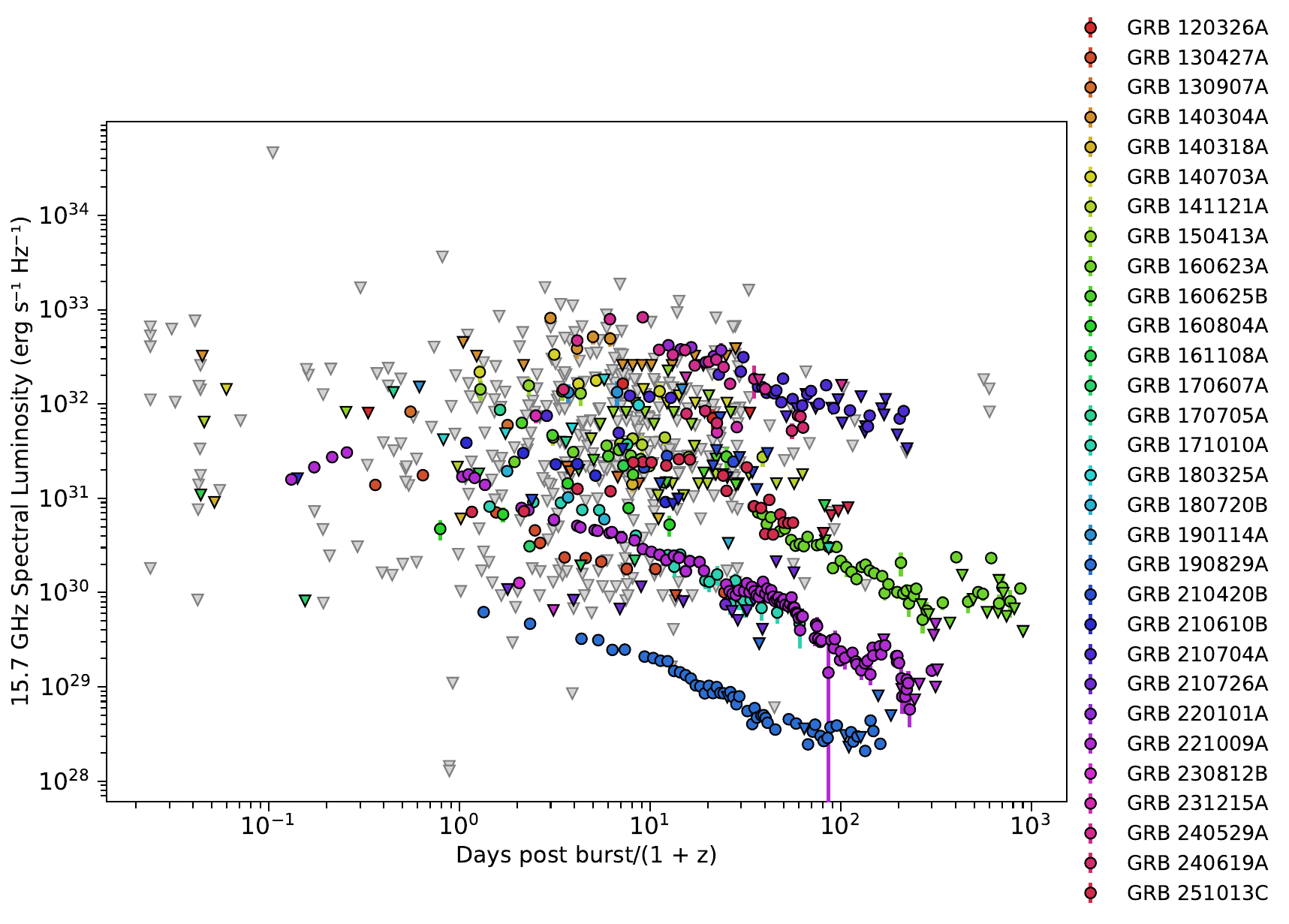}
    \caption{The \textit{k}-corrected spectral luminosity light curves of the 103 AMI-detected GRBs in the rest frame from both AMI-GRB-1 and 2 including 30 detections \citep{2018MNRAS.473.1512A}. The GRBs are color coded for distinction, with their detections represented as dots and their 3$\sigma$ upper-limits represented as inverse triangles. The GRBs with a redshift measurement which were observed but not-detected by AMI are represented by gray triangles.}
    \label{fig:Anderson Redshift Lightcurves}
\end{figure}

\begin{figure}[h!]
\begin{subfigure}{\linewidth}
  \includegraphics[width=\linewidth]{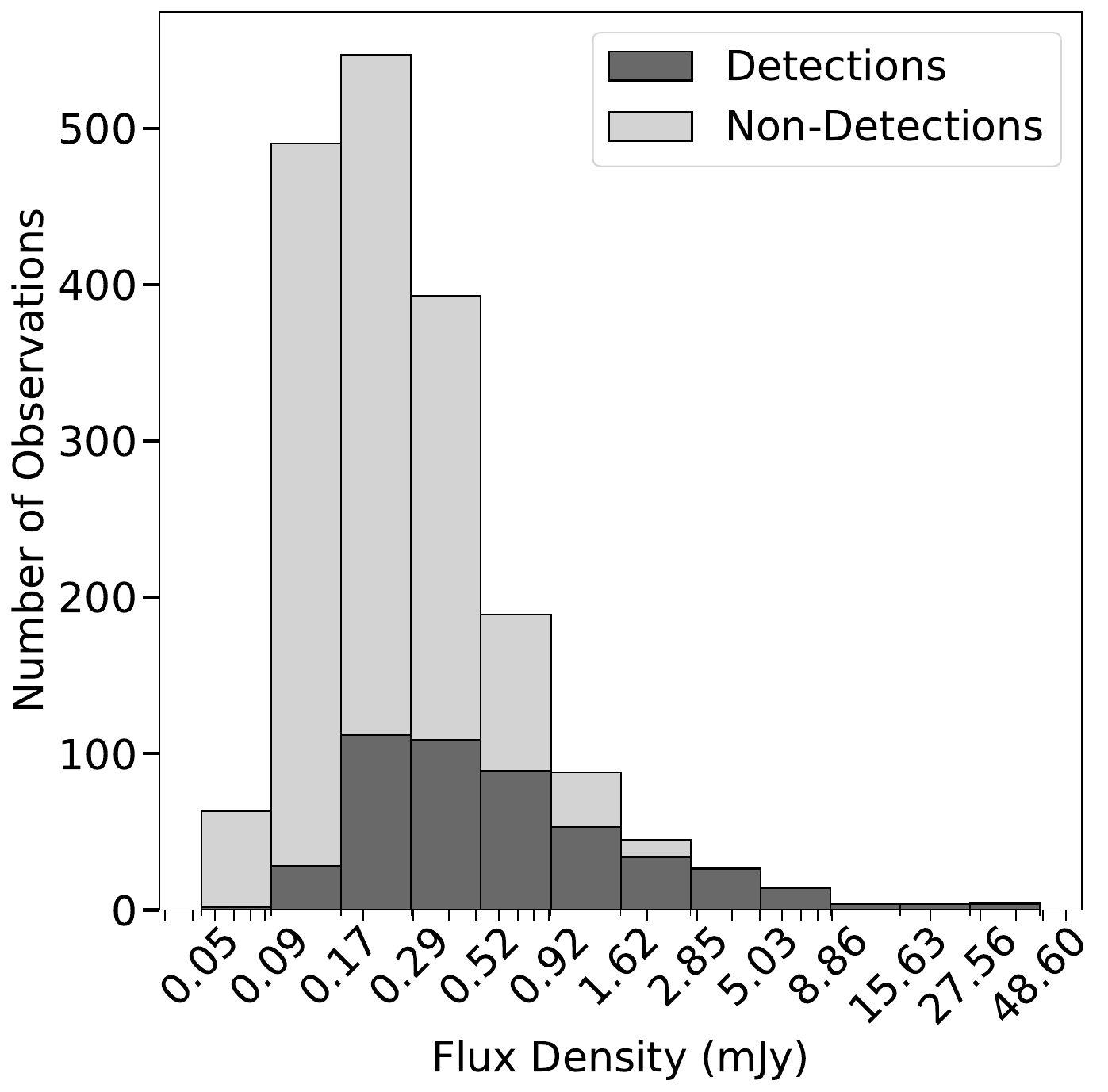}
  \caption{Distribution of entire population data.}
  \label{fig: Anderson Flux Density All}
\end{subfigure}\hfill % <-- "\hfill"
~ % optional tilde b/t figures for readability.
  % this solution will not work w/ empty lines b/t subfigures
\begin{subfigure}{\linewidth}
  \includegraphics[width=\linewidth]{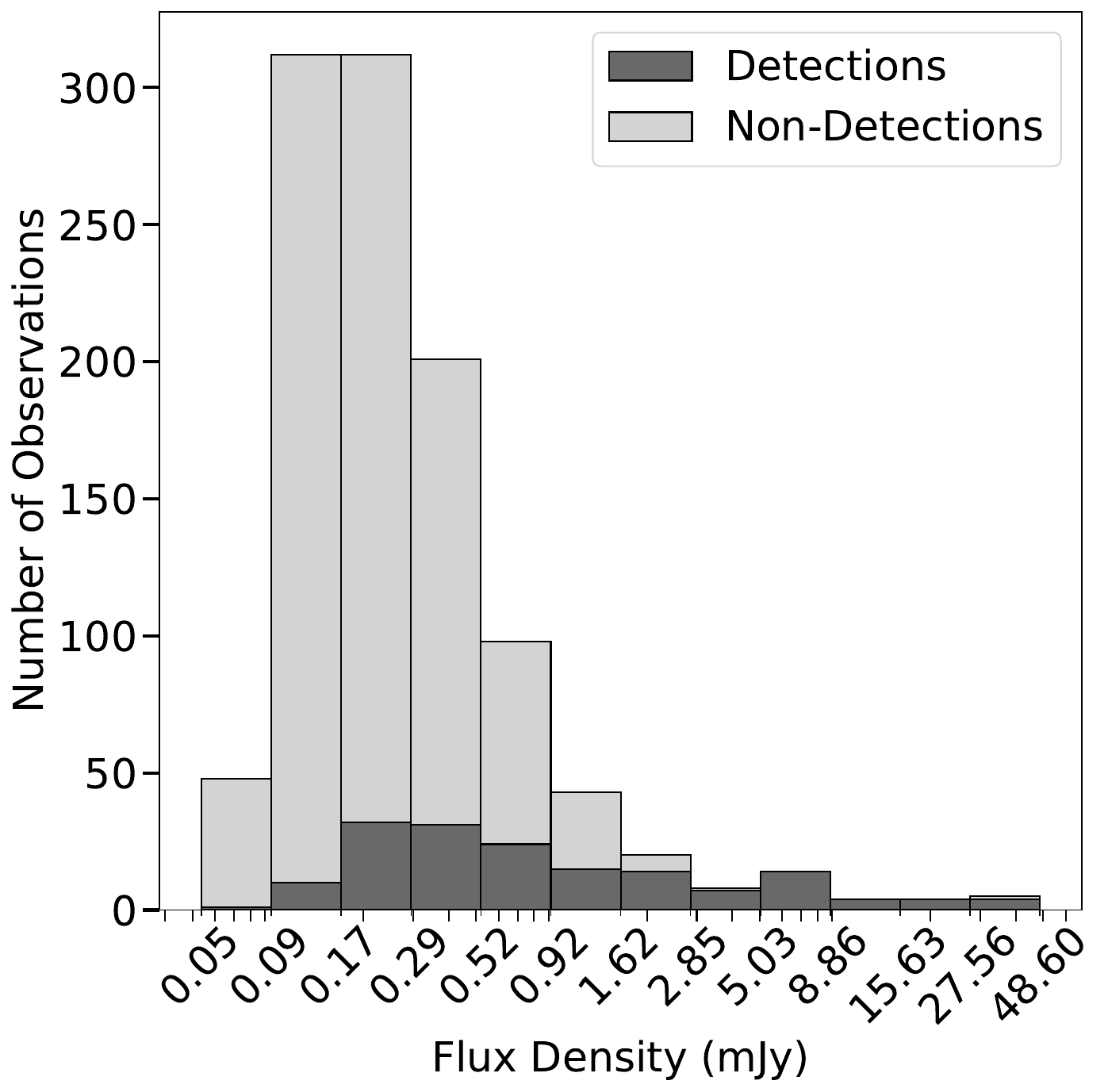}
  \caption{Distribution of data with $t \le 10$ days.}
  \label{fig: Anderson Flux Density 10 days}
\end{subfigure}
\label{fig: Flux density dist Anderson}
\caption{Stacked histograms showing the flux density distribution of all observations within AMI-GRB-1 and 2 (top) and of the observations between 0 and 10 days since burst  \citep[bottom, ][]{2018MNRAS.473.1512A} The light gray bars represent the distribution for the 3$\sigma$ upper-limits, and the dark gray bars represent the distribution of the detections. The early-time distribution of detections is flat across the range of flux densities. The typical 3$\sigma$ upper limit of any AMI-LA observation is 0.1-0.2\,mJy.}
\end{figure}

\begin{figure}%[19]{0.45\textwidth}
    \centering
    \includegraphics[width=\linewidth ]{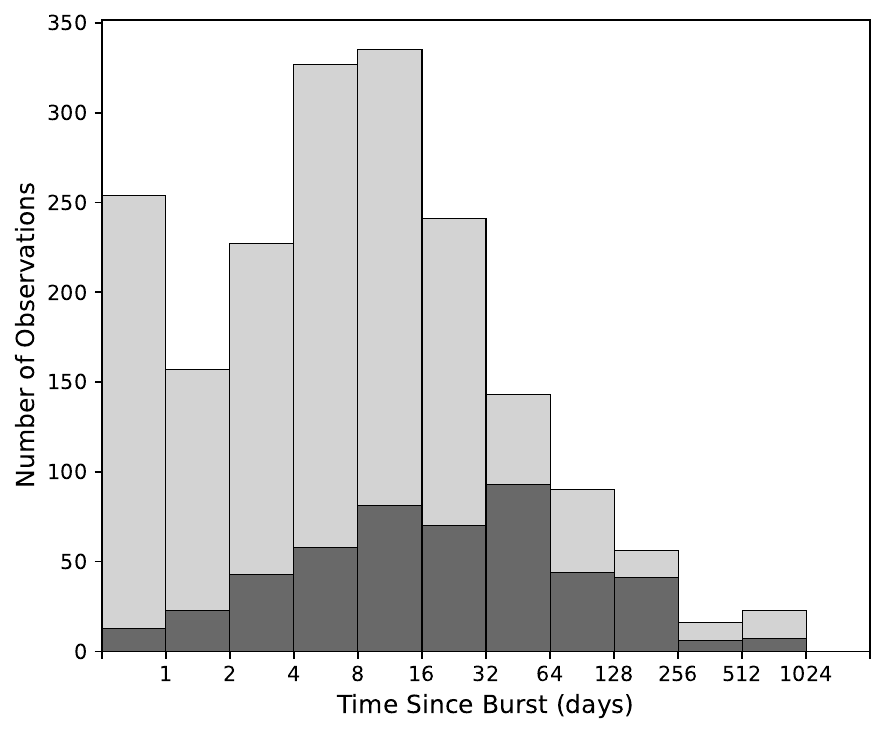}
    \caption{The distribution of observing times post-GRB for both AMI-GRB-1 and AMI-GRB-2 \citep{2018MNRAS.473.1512A}.}
    \label{fig: Time dist Anderson}
\end{figure}

\begin{figure}%[19]{0.45\textwidth}
    \centering
        \includegraphics[width=\linewidth ]{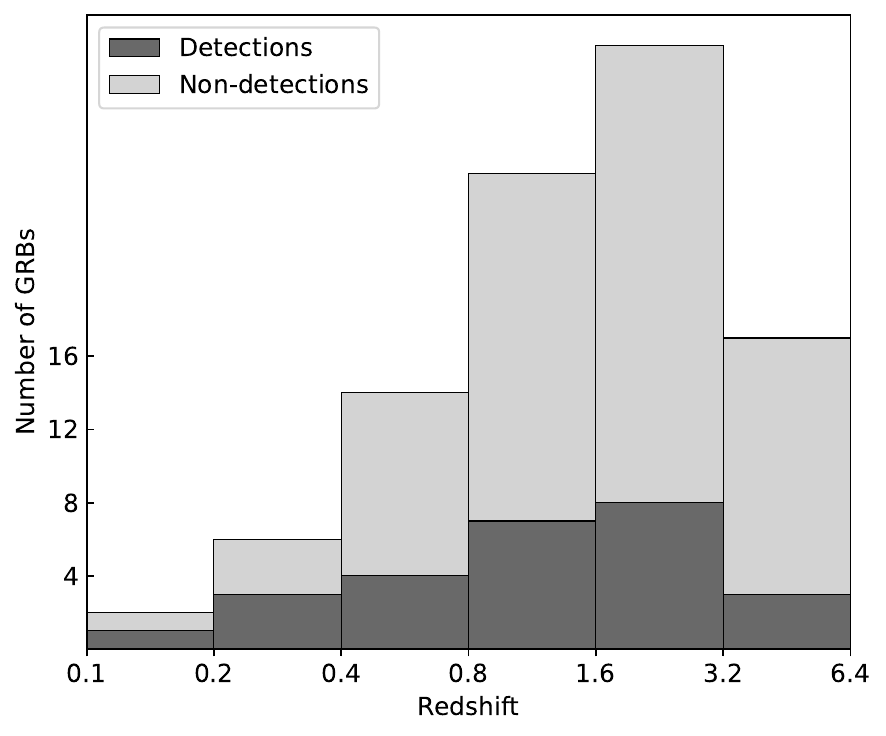}
    \caption{Redshift distribution of the 100 GRBs with redshift measurements from both AMI-GRB-1 and AMI-GRB-2 \citep{2018MNRAS.473.1512A}.}
    \label{fig: Anderson Redshift Distribution}
\end{figure}

\newpage
\newpage
\section{Observation table}

%% For this sample we use BibTeX plus aasjournalv7.bst to generate the
%% the bibliography. The sample7.bib file was populated from ADS. To
%% get the citations to show in the compiled file do the following:
%%

\begin{table*}[!ht]
    \centering
    \caption{Data table for AMI-GRB-2 consisting of the GRB name, the date and central time of each observation in MJD, the peak flux density in mJy and the rms noise in the image (also in mJy) or in the case of a non-detection the 3$\sigma$ upper limit. The full table will be available in a machine readable format. }
    \begin{tabularx}{0.85\textwidth}{cccc}
    \hline
    \hline
        GRB Name & Observation Date (MJD) & Peak flux density (mJy) & Uncertainty/3$\sigma$ upper limit (mJy)\\
    \hline
GRB 160131A & 57445.205 & - & 0.180\\
GRB 160227A & 57450.238 & - & 0.177\\
GRB 160227A & 57454.097 & - & 0.114\\
GRB 160227A & 57459.177 & - & 0.102\\
GRB 160227A & 57514.145 & - & 0.105\\
GRB 160227A & 57560.95 & - & 0.099\\
GRB 160227A & 57563.918 & - & 0.183\\
GRB 160313A & 57460.153 & - & 0.117\\
GRB 160313A & 57460.976 & - & 0.066\\
GRB 160313A & 57463.001 & - & 0.093\\
GRB 160313A & 57467.118 & - & 0.174\\
GRB 160313A & 57470.03 & - & 0.111\\
GRB 160313A & 57545.795 & - & 0.201\\
.. & .. & .. & ..\\
    \hline
    \hline
    \end{tabularx}

\end{table*}
%\bibliography{sample701}{}
%\bibliographystyle{aasjournalv7}

%% This command is needed to show the entire author+affiliation list when
%% the collaboration and author truncation commands are used.  It has to
%% go at the end of the manuscript.
%\allauthors

%% Include this line if you are using the \added, \replaced, \deleted
%% commands to see a summary list of all changes at the end of the article.
%\listofchanges

\end{document}